\documentclass[12pt]{article}

\usepackage{subfigure}
\usepackage{amssymb}
\usepackage{amsfonts}
\usepackage{bm}
\usepackage{color}
\usepackage{calc}
\usepackage{amsfonts}
\usepackage{amsmath}
\usepackage[english]{babel}
\usepackage[T1]{fontenc}
\usepackage{verbatim}
\usepackage{subfig}
\usepackage{enumerate}
\usepackage{graphicx}
\usepackage{natbib}
\usepackage{url}
\usepackage{algorithmic,algorithm}
\usepackage{placeins}

\def\D{\partial}

\def\I{{\mathcal{I}}}

\def\C{{\mathcal{C}}}
\def\A{{\mathcal{A}}}

\def\N{{\mathcal{N}}}
\def\P{{\mathcal{P}}}
\def\Q{{\mathcal{Q}}}

\def\L{{\mathcal{L}}}
\def\D{{\mathcal{D}}}
\def\G{{\mathcal{G}}}
\def\T{{\mathcal{T}}}

\def\RR{{\mathbb{R}}}
\def\CC{{\mathbb{C}}}

\def\ZZ{{\mathbb{Z}}}
\def\bA{{\bf A}}
\def\bC{{\bf C}}
\def\bD{{\bf D}}
\def\bE{{\bf E}}
\def\bG{{\bf G}}
\def\bF{{\bf F}}

\def\bB{{\bf B}}

\def\bK{{\bf K}}
\def\bM{{\bf M}}

\def\DK{{\bf DK}}

\def\e{{\mathbf e}}

\def\n{{\mathbf n}}
\def\x{{\bm x}}

\def\t{{\mathbf t}}
\def\c{{\mathbf c}}

\def\x{{\mathbf x}}
\def\y{{\mathbf y}}
\def\z{{\mathbf z}}

\def\v{{\mathbf v}}
\def\w{{\mathbf w}}

\def\0{{\mathbf 0}}

\def\bnabla{\boldsymbol{\nabla}}
\def\bDelta{\boldsymbol{\Delta}}

\def\bXi{\boldsymbol{\Xi}}

\def\Dpartial#1#2{ {\partial #1 \over \partial #2} }

\def\Bmp#1{ \begin{minipage}{#1} }
\def\Emp{ \end{minipage} }
\def\Bmpc#1{ \begin{minipage}[c]{#1} }
\def\Bmpt#1{ \begin{minipage}[t]{#1} }
\def\Bmpb#1{ \begin{minipage}[b]{#1} }

\newcommand{\diag}{\operatorname{diag}}
\newcommand{\Sp}{\operatorname{sp}}

\begin{document}
\title{Linear Stability of Inviscid Vortex Rings to Axisymmetric Perturbations}

\author{Bartosz Protas\thanks{Email address for correspondence: bprotas@mcmaster.ca} 
\\ \\ 
Department of Mathematics and Statistics, McMaster University \\
Hamilton, Ontario, L8S 4K1, Canada
}

\date{\today}

\maketitle

\begin{abstract}
  We consider the linear stability to axisymmetric perturbations of
  the family of inviscid vortex rings discovered by
  \citet{norbury-1973-JFM}.  Since these vortex rings are obtained as
  solutions to a free-boundary problem, their stability analysis is
  performed using recently-developed methods of shape differentiation
  applied to the contour-dynamics formulation of the problem in the 3D
  axisymmetric geometry. This approach allows us to systematically
  account for the effects of boundary deformations on the linearized
  evolution of the vortex ring. We investigate the instantaneous
  amplification of perturbations assumed to have the same the
  circulation as the vortex rings in their equilibrium configuration.
  These stability properties are then determined by the spectrum of a
  singular integro-differential operator defined on the vortex
  boundary in the meridional plane. The resulting generalized
  eigenvalue problem is solved numerically with a spectrally-accurate
  discretization. Our results reveal that while thin vortex rings
  remain neutrally stable to axisymmetric perturbations, they become
  linearly unstable to such perturbations when they are sufficiently
  ``fat''. Analysis of the structure of the eigenmodes demonstrates
  that they approach the corresponding eigenmodes of Rankine's vortex
  and Hill's vortex in the thin-vortex and fat-vortex limit,
  respectively. This study is a stepping stone on the way towards a
  complete stability analysis of inviscid vortex rings with respect to
  general perturbations.
\end{abstract}

\begin{flushleft}
\end{flushleft}



\section{Introduction}
\label{sec:intro}

Vortex rings are among the most ubiquitous vortex structures
encountered in fluid mechanics, both in laminar and turbulent flows.
They occur in flow phenomena arising in diverse situations such as
aquatic propulsion \citep{dabiri-2009-AR}, blood flow in the human
heart \citep{kheradvar-2012,Arvidsson2016} and in detonations
\citep{Giannuzzi2016}, to mention just a few. A typical situation
leading to the formation of {``thin''} vortex rings is when a jet
issues from an orifice \citep{maxworthy_1977}, which is also the
design principle of the ``vortex gun'' \citep{vortex_gun}.  {On
  the other hand, ``fat'' vortex rings have been invoked as models
  of steady wakes behind spherical objects \citep{gmgdjw08}.}  Under
certain simplifying assumptions vortex rings are known to represent
relative equilibria of the equations governing the fluid motion and
the persistence of these states is determined by their stability
properties. The goal of this investigation is to address this question
in the context of an idealized inviscid model of a vortex ring.

Given their significance in nature and technology, vortex rings have
received a lot of attention in the fluids-mechanics literature, both
from the experimental and theoretical point of view
\citep{akhmetov-2009}. However, since in the general setting of
three-dimensional (3D) unsteady flows the Navier-Stokes system is
analytically intractable, most investigations in such settings have
been based on numerical studies. Analytical approaches are
applicable subject to some simplifying assumptions, usually involving
axisymmetry and steady state. Starting from the original work of
\citet{tung-1967}, the effects of viscosity were considered by
\citet{saffman-1970,bk92a,fukumoto-2000,kaplanski-2005-PF,fukumoto-2008}
among others, in most cases based on techniques of asymptotic
analysis.

Significant advances have been made in the study of vortex rings in
the inviscid limit, where some of the key results go back to the
seminal works of by \citet{Kelvin1867} and \citet{Hicks1899}.
Existence of inviscid vortex rings of finite thickness and with
nontrivial core structure was predicted theoretically by
\citet{Fraenkel1970} and \citet{norbury-1972-proc}. Then,
\citet{norbury-1973-JFM} computed a one-parameter family of vortex
rings which contains the infinitely thin vortex ring and Hill's
spherical vortex \citep{hill-1894} as the limiting members. Rigorous
interpretation of such solutions in terms of variational principles
was later developed by \citet{Benjamin1975,Wan1988} and
\citet{fukumoto-2008-PhysD}.  For an additional discussion of vortex
rings the reader is referred to the monographs by
\citet{saffman-1992,lamb-1932,wu-book-2006,alekseenko2007theory}. The
principal objective of the present investigation is to shed light on
the stability of Norbury's family of inviscid vortex rings to
axisymmetric perturbations.

As regards the stability analysis of vortex rings, to date research
has been primarily focused on two cases representing the limiting
members of Norbury's family, namely, thin vortex rings and Hill's
spherical vortex, the latter of which can be regarded as a limiting
``fat'' vortex ring.  {In both cases the analysis has focused on
  temporal growth of perturbations.}  Concerning the former case, thin
vortex rings were treated as a vortex filament embedded in the ring's
own straining field {represented as a quadrupole}, a
configuration susceptible to the so-called Moore-Saffman-Tsai-Widnall
instability
{\citep{WidnallSullivan1973,widnall_bliss_tsai_1974,MooreSaffman1975,WidnallTsaiStuart1977},
  which is an example of a short-wavelength instability
  \citep{wu-book-2006} and results in a wavy bending deformation of
  the vortex ring}.  These results were later extended by
\citet{HattoriFukumoto2003} and \citet{fukumoto_hattori_2005} who
identified a different instability mechanism {in which curvature
  effects represented as a local dipole field result in the stretching
  of the perturbed vorticity in the toroidal direction.}  Since the
vortex rings are assumed thin, and hence are locally approximated as
Rankine's columnar vortex, in this analysis the deformation of the
interface between the vortex core and the irrotational flow has the
form of Kelvin waves. {The response of Norbury's vortex rings of
  various thickness to prolate and oblate axisymmetric perturbations
  of their shape was first studied numerically by \citet{YeChu1995}
  and then more systematically by \citet{Ofarrell_dabiri_2012}. Among
  other observations, these investigations demonstrated that when they
  are sufficiently fat, perturbed axisymmetric vortex rings shed
  vorticity in the form of thin sheets.}  On the other hand, in the
case of Hill's spherical vortex the instability has the form of a
sharp deformation of the vortex boundary localized at the rear
stagnation point \citep{mm78}.  {The response of Hill's vortex to
  general, non-axisymmetric perturbations was studied using
  approximate techniques based on expansions of the flow variables in
  spherical harmonics and numerical integration by \citet{frk94,r99}.
  Their main findings were consistent with those of \citet{mm78},
  namely, that perturbations of the vortex boundary develop into sharp
  spikes whose number depends on the azimuthal wavenumber of the
  perturbation.  A number of investigations \citep{l95,rf00,hh10}
  studied the stability of Hill's vortex with respect to
  short-wavelength perturbations applied locally and advected by the
  flow in the spirit of the WKB approach \citep{lh91}. These analyses
  revealed the presence of a number of instability mechanisms,
  although they are restricted to the short-wavelength regime. In this
  context we also mention the study by \citet{lsf01} who considered
  the linear response of the compressible Hill's vortex to acoustic
  waves.}

The study of the stability of axisymmetric vortex rings has several
analogies to the study of the stability of equilibrium configurations
of two-dimensional (2D) vortex patches which has been investigated
extensively using both analytical techniques \citep{k80,l93,b90,ghs04}
and computational approaches \citep{d85,k87,d90,dl91,d95,efm05}.  In
the former context we also mention the recent study by \citet{gs18}
who investigated the spectral stability of inviscid columnar vortices
with respect to general 3D perturbations.  Most importantly, the task
of finding equilibrium configurations (of axisymmetric vortex rings in
3D or vortex patches in 2D) is a {\em free-boundary} problem, because
the shape of the boundary separating the vortical and irrotational
flows is a priori unknown and must be found as a part of the solution
to the problem. Studying the stability of such configurations is in
general situations a non-trivial task, because instabilities involve
deformations of the boundaries which must be suitably parameterized in
order to perform linearization. A general approach to handle such
problems was recently developed based on methods of differential
geometry, known as the shape calculus \citep{dz01a}, by \citet{ep13}.
{An important property of this approach is the fact that the
  stability analysis, including suitable linearization, is performed
  on the continuous problem before any discretization is applied.}
This approach {thus} provides a versatile framework to study the
stability of general vortex equilibria and the classical results
concerning the stability of Rankine's and Kirchhoff's vortices due to
\citet{k80} and \citet{l93} can be derived from it as special cases
through analytical computations \citep{ep13}.  Using this method the
authors performed a complete linear stability analysis of Hill's
vortex with respect to axisymmetric perturbations
\citep{ProtasElcrat2016}, which complemented and refined the original
findings of \citet{mm78}. This analysis revealed the presence of two
linearly unstable eigenmodes in the form of singular distributions
localized at the rear stagnation point and a continuous spectrum of
highly oscillatory neutrally-stable modes.

As the main contribution of the present study, we use the approach of
\citet{ep13} and \citet{ProtasElcrat2016} to carry out a linear
stability analysis of Norbury's family of vortex rings with respect to
axisymmetric perturbations. {More precisely, we investigate
  whether or not perturbations to the equilibrium shapes of the vortex
  boundary can be instantaneously amplified in time.}  This analysis,
performed for vortex rings with varying ``fatness'', is a stepping
stone towards obtaining a complete picture of the stability of
inviscid vortex rings with respect to arbitrary perturbations. Our
analysis demonstrates that while thin vortex rings are neutrally
stable, they become linearly unstable once they are sufficiently
``fat''. Careful analysis of the spectra and of the structure of the
eigenmodes reveals that, as expected, they approach the spectra and
the corresponding eigenvectors of Rankine's vortex and Hill's vortex
as the vortex ring becomes, respectively, thin or fat. {We
  emphasize that the validity of these results is limited by the
  assumption of axisymmetry imposed on the perturbations and when
  perturbations of a general form are allowed, then thin vortex rings
  have also been found to be unstable (cf.~the discussion above).}

{In comparison to our earlier study of the stability of Hill's
  vortex \citep{ProtasElcrat2016}, now the main complication is that
  the equilibrium shapes of the vortex boundary are no longer
  available in a closed form and must be determined numerically for
  different members of Norbury's family. While our approach to the
  stability analysis is essentially the same as used by
  \citet{ProtasElcrat2016}, in that study the problem was more
  complicated due to the degeneracy of the stability equation at the
  two stagnation points which necessitated special regularization. On
  the other hand, for Norbury's vortex rings the stability problem is
  in principle well-posed, but its numerical conditioning
  significantly deteriorates for vortex rings with increasing fatness,
  i.e., when Norbury's vortices approach Hill's spherical vortex.}

The structure of the paper is as follows: in the next section we
introduce Norbury's family of inviscid vortex rings as solutions to
steady Euler's equations in the 3D axisymmetric geometry and recompute
the solutions originally found by \citet{norbury-1973-JFM} with much
higher accuracy; next, in \S \ref{sec:stab}, we present a linear
stability analysis based on shape differentiation and obtain a
generalized eigenvalue problem for a singular integro-differential
operator whose solutions encode information about the stability of the
vortex rings; a spectrally-accurate approach to numerical solution of
this generalized eigenvalue problem is described in \S
\ref{sec:numer}, whereas our computational results are presented in \S
\ref{sec:comput}; discussion and conclusions are deferred to \S
\ref{sec:final}; some technical details are collected in an appendix.

\section{The Family of Norbury's Vortex Rings}
\label{sec:rings}

In this section we recall the family of inviscid vortex rings
discovered by \citet{norbury-1973-JFM}. Since these flows are
axisymmetric, we will adopt the cylindrical-polar coordinate system
$(r,\phi,z)$ and define the operator $\L :=
\bnabla\cdot\left(\frac{1}{r}\, \bnabla\right)$ in which $\bnabla
:= \left[\Dpartial{}{r}, \Dpartial{}{z}\right]^T$ (``:='' means
``equal to by definition''). Given the Stokes streamfunction $\psi =
\psi(r,z)$, these flows satisfy the following system in the frame
of reference moving in the direction of the $z$ axis with the
translation velocity $W$ of the vortex
\begin{subequations}
\label{eq:Euler3D}
\begin{alignat}{2}
\L \psi & = - r \, f(\psi) 
&\quad & \textrm{in} \ \Omega, \label{eq:Euler2Da} \\
\psi & \rightarrow  -\frac{1}{2}W r^2 &  & \textrm{as} \ |\x| := \sqrt{r^2+z^2} \rightarrow \infty,
\label{eq:Euler2Db} 
\end{alignat}
\end{subequations}
where {$\Omega = \RR^2$ is the meridional plane,} $\x = [r, z]^T$
and the vorticity function $f(\psi)$ has the form
\begin{equation}
f(\psi) = \left\{
\begin{tabular}{ll}
$\C$, \quad &  $\psi > {\psi_0}$\\
0,   \quad & $\psi \le {\psi_0}$
\end{tabular}
\right.,
\label{eq:f}
\end{equation}
in which $\C \neq 0$ and ${\psi_0} \ge 0$ are constants. System
\eqref{eq:Euler3D}--\eqref{eq:f} therefore describes a compact region
$\D$ with azimuthal vorticity $\omega := \omega_{\phi}$ {varying}
proportionally to the distance $r$ from the flow axis embedded in a
potential flow. The boundary of this region $\partial \D := \left\{
  (r,\phi,z)\; : \; \psi(r,z) = {\psi_0} \right\}$ is a priori
unknown and must be found as a part of the problem solution. System
\eqref{eq:Euler3D}--\eqref{eq:f} thus represents a {\em free-boundary}
problem and, as will become evident below, this property makes the
study of the stability of its solutions more complicated.

\begin{figure}
\centering
\includegraphics[width=0.5\textwidth]{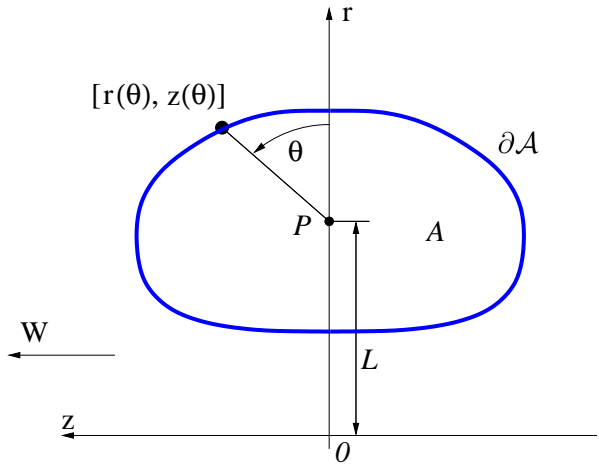}
\caption{Schematic representation of the problem geometry in the
  meridional plane $(r,z)$ {with a representative member of
    Norbury's family of vortex rings (marked with a think blue
    line)}.}
\label{fig:A}
\end{figure}

Solution of the stability problem requires a convenient representation
of the boundary $\partial\D$ of the vortex ring. The intersection of
the vortex region $\D$ with the meridional plane $(r,z)$, which we
will denote $\A$, is shown schematically in figure \ref{fig:A}. We
consider the following general parametrization of the vortex boundary
$\x(\theta) = [r(\theta),z(\theta)]^T$, where $\theta \in [0,2\pi]$
{and $r \; : \; [0,2\pi] \rightarrow \RR^+$, $z \; : \;[0,2\pi]
  \rightarrow \RR$} are smooth periodic functions, which can be viewed
as a mapping from a unit circle to $\partial\A$. In his original work,
\citet{norbury-1973-JFM} employed the following particular form of
this parametrization {adapted to solutions with the fore-and-aft
  symmetry and} given in terms of {some even} periodic function
$g \; : \; [0,2\pi] \rightarrow \RR$
\begin{subequations}
\label{eq:rz}
\begin{align}
r(\theta) &= g(\theta) \cos\theta + 1,      \label{eq:rr} \\
z(\theta) &= g(\theta) \sin\theta.      \label{eq:z}
\end{align}
\end{subequations}
However, in our study we retain the more general form of the
parametrization expressed in terms of two functions $r = r(\theta)$
and $z = z(\theta)$, {a choice motivated by the fact that in our
  subsequent stability analysis we will consider boundary
  perturbations which break the fore-and-aft symmetry of the
  equilibrium configurations (in principle, such a more general
  representation of the vortex boundary $\partial \A$ could allow one
  to also capture equilibrium solutions without the fore-and-aft
  symmetry, but no such solutions were found).}  The point $P$ is
defined as lying halfway between the two radial extremities of the
vortex region $\A$, i.e., we have $|OP| =: L =
{(1/2)[r(0)+r(\pi)]}$, and, following \citet{norbury-1973-JFM},
we will normalize the problem such that $L=1$.  Expressing the area of
the vortex region as $|\A| = \pi L^2 \alpha^2 = \pi \alpha^2$ allows
us to define the mean core radius $\alpha$.  Assuming {a
  particular reference velocity $U$,} the constant in \eqref{eq:f} can
be expressed as {$\C = U / (L^2\alpha^2)$, such that after
  non-dimensionalizing the problem and noting the normalization
  mentioned above, this constant becomes $\C = 1 / \alpha^2$
  {\citep{norbury-1973-JFM}}. This leaves} $\alpha$ as the sole
quantity parameterizing the solutions of system
\eqref{eq:Euler3D}--\eqref{eq:f}, and thus also implicitly determining
$W$ and {$\psi_0$}. We note that as $\alpha \rightarrow 0$ the
solutions of \eqref{eq:Euler3D}--\eqref{eq:f} approach an
infinitesimally thin vortex ring, whereas for $\alpha \rightarrow
\sqrt{2}$ they approach Hill's vortex in which the region $\A$ becomes
a half-disk of radius 2.

A consistent way to study the evolution and equilibria of inviscid
flows with discontinuous vorticity distributions is based on the
formalism of contour dynamics \citep{p92}. Given a time-dependent
region $\A(t)$, where $t$ is time, its evolution can be studied by
tracking the points $\y(t)$ on its boundary $\partial\A(t)$ via the
equation
\begin{equation}
\frac{d\y(t)}{dt} = \v(\y(t)) := \C \, \int_{\partial\A(t)} \bK(\y(t),\y')\, ds_{\y'}, \quad
\forall \y(t) \in \partial\A(t),
\label{eq:v}
\end{equation}
where $\v$ denotes the velocity, $\bK(\y(t),\y')$ is a suitable
Biot-Savart kernel, $\y$ and $\y'$ are defined in the absolute frame
of reference and $ds_{\y'}$ is an arclength element of the vortex
boundary in the meridional plane.  An equilibrium shape of the vortex
boundary $\partial\A$ when the right-hand side (RHS) in
\eqref{eq:Euler2Da} has a step discontinuity, cf.~\eqref{eq:f}, can be
characterized by transforming the coordinates to the translating frame
of reference $\x(t) = \y(t) - W t\, \e_z$ and considering the normal
component of equation \eqref{eq:v}
\begin{equation}
\n_{\x}\cdot\frac{d\x(t)}{dt} =  \n_{\x}\cdot\left[
\C \, \int_{\partial\A} \bK(\x(t),\x')\, ds_{\x'} - W\, \e_z\right] = 0, 
\label{eq:vn}
\end{equation}
where $\n_{\x} := [n_r, n_z]^T$ denotes the unit vector normal to the
contour $\partial\A$ at the point $\x$ and directed outward.
{Equation \eqref{eq:vn} expresses the vanishing of the normal
  velocity component on the vortex boundary in relative equilibrium.}
Further below we will also use the unit tangent vector $\t_{\x} :=
[t_r, t_z]^T$ at the point $\x$ (we will adhere to the convention that
a subscript on a geometric quantity, such as the tangent or normal
vector and the curvature, will indicate where this quantity is
evaluated). Since we have $\x(0) = \y(0)$, the arguments of the kernel
$\bK$ could be changed {from $\y$ and $\y'$ in \eqref{eq:v} to
  $\x$ and $\x'$ in \eqref{eq:vn}}.  {As} the equilibrium shape
of the boundary $\partial\A$ is in general a priori unknown, relation
\eqref{eq:vn} reveals the free-boundary aspect of the problem. The
Biot-Savart kernel {$\bK$ appearing in \eqref{eq:v} and
  \eqref{eq:vn} is obtained from the standard form of this kernel in
  3D by integrating it with respect to the azimuthal angle $\phi$,
  which is justified by the axisymmetry assumption. As a result, in
  relations \eqref{eq:v}--\eqref{eq:vn} integration needs to be
  performed in the meridional plane only. Expressions for such an
  axisymmetric Biot-Savart kernel were} derived by {\citet{wr96}} with
an alternative, but equivalent, formulation also given by \citet{p86}.
{Here we will use this kernel in the form rederived and tested by
  \citet{slf08}}
\begin{equation}
\begin{aligned}
\bK(r\,\e_r+z\,\e_z, r'\,\e_r+z'\,\e_z) =
& -r'  H(r,z,r',z') \,  n_z(r',z') \, \e_{r}  \\
& \hspace*{-2.5cm} -\left[ (z'-z) \, G(r,z,r',z') \, n_z(r',z') -  
r \, H(r,z,r',z') \,  n_{r}(r',z') \right]\, \e_z,
\end{aligned}
\label{eq:K}
\end{equation}
where 
\begin{align*}
G(r,z,r',z') &:=  \frac{r'}{\pi \sqrt{A+B}} K(\sigma), \\
H(r,z,r',z') &:=  \frac{1}{2\pi r} \left[ \frac{A}{\sqrt{A+B}} K(\sigma) - E(\sigma)\, \sqrt{A+B}\right]
\end{align*}
in which 
\begin{equation*}
A :=  r^2 + r'^2 + (z-z')^2, \qquad B := 2 rr', \qquad \sigma := \sqrt{\frac{2B}{A+B}},
\end{equation*}
whereas $K(\sigma)$ and $E(\sigma)$ are the complete elliptic
integrals of the first and second kind, respectively \citep{olbc10}.
We note that $\sigma \rightarrow 1$ as $r' \rightarrow r$ and $z'
\rightarrow z$, and at $\sigma=1$ the function $K(\sigma)$ has a
logarithmic singularity (more details about the singularity structure
of kernel \eqref{eq:K} will be provided in \S \ref{sec:stab}).

Combining the kinematic condition \eqref{eq:vn} with the normalization
condition {$L = 1$}, we obtain the following system of equations
defining solutions of problem \eqref{eq:Euler3D}--\eqref{eq:f}, i.e.,
the vortex region $\A$ and the translation velocity $W$, {as a}
function of the parameter {(the mean core radius)} $\alpha \in
(0,\sqrt{2}]$
\begin{subequations}
\label{eq:CD}
\begin{align}
\frac{1}{\alpha^2} \n_{\x}\cdot\int_{\partial\A} \bK(\x,\x')\, ds_{\x'} 
- W\, \n_{\x}\cdot\e_z  & = 0, {\qquad \forall \x \in \partial\A} \label{eq:CDa} \\
{r(0)+r(\pi)} - 2 & = 0, \label{eq:CDb} \\
|\A| - \pi \alpha^2 & = 0. \label{eq:CDc}
\end{align}
\end{subequations}
In order to obtain his family of vortex rings,
\citet{norbury-1973-JFM} solved a system equivalent to \eqref{eq:CD}
in which equation \eqref{eq:CDa} was replaced with the condition
{$\psi = \psi_0$} on $\partial\A$, where the streamfunction
$\psi$ was evaluated using an integral expression involving Green's
function of the operator $\L$. He did this by representing the
function $g$ in \eqref{eq:rz} in terms of a truncated cosine series
and then solving the resulting discrete system using Newton's method
in which the entries of the Jacobian matrix were approximated using
finite differences.  The results are reported in
\citet{norbury-1973-JFM} in the form of cosine-series coefficients of
the function $g$, cf.~\eqref{eq:rz}, obtained for different values of
$\alpha \in (0,\sqrt{2})$. While in principle one could use these
solutions to define the base states in our stability analysis, we will
recompute the family of vortex rings to much higher precision using a
method which is more accurate than Norbury's original approach. In
addition, this will also allow us to find members of this family for
values of the parameter $\alpha$ not considered by
\citet{norbury-1973-JFM}, which will be important for understanding
exactly how the stability properties of the vortex rings vary with
$\alpha$, i.e., in function of the ``fatness'' of the vortex rings. It
is preferable to solve the problem based on formulation \eqref{eq:CD},
rather than the one used originally by \citet{norbury-1973-JFM},
because in the former case the Jacobian computed with respect to the
shape of the contour $\partial\A$ can be reused in a straightforward
manner for the stability calculations.  Below we present a brief
outline of this approach.

{The idea of the approach is to find a deformation of some
  initial reference contour such that conditions
  \eqref{eq:CDa}--\eqref{eq:CDc} are satisfied. System \eqref{eq:CD}
  is solved, in a suitable discretized form, using Newton's method in
  which the required contour deformation (Newton's step) is found
  approximately by solving a linear system (Jacobian) obtained by
  differentiating \eqref{eq:CD} with respect to the shape of the
  contour $\partial \A$. Taking $\x$ to represent the contour
  approximating the vortex boundary $\partial \A$ at a certain
  iteration, its perturbation $\x^{\epsilon}$ obtained by applying a}
small but otherwise arbitrary deformation can be described as
\begin{equation}
\x^{\epsilon}(\theta) = \x(\theta) + \epsilon\, \rho(\theta) \, \n_{\x(\theta)},
\label{eq:xeps}
\end{equation}
where $\epsilon > 0$ and $\rho \; : \; [0,2\pi] \rightarrow \RR$ is a
periodic function describing the shape of the perturbation. We will
adopt the convention that the superscript $\epsilon$ will denote
quantities corresponding to the perturbed boundary, so that $\x =
\x^{\epsilon}|_{\epsilon = 0}$ and $\n_{\x} =
\n_{\x^{\epsilon}}|_{\epsilon = 0}$ are the quantities corresponding
to the reference shape.  Since an arbitrary contour $\partial\A$ does
not represent an equilibrium, {system} \eqref{eq:CD} {defining
  the equilibrium} can be equivalently expressed as the following
conditions on the deformation $\rho$ and the translation velocity $W$
{obtained by setting $\epsilon = 1$ in ansatz \eqref{eq:xeps}}
\begin{equation}
F\left(\begin{bmatrix} \rho \\ W \end{bmatrix}\right)
:= \begin{bmatrix} 
{\n_{\x(\theta) + \rho(\theta) \, \n_{\x(\theta)}}\cdot \v(\x(\theta) + \rho(\theta) \, \n_{\x(\theta)})- W\, \n_{\x(\theta) + \rho(\theta) \, \n_{\x(\theta)}}\cdot\e_z} \\
{\left[r(0) + \rho(0)\, n_r(0)\right] + \left[ r(\pi)+ \rho(\pi)\, n_r(\pi)\right] - 2} \\
{|\A(\rho)|} - \pi \alpha^2
\end{bmatrix}
= \begin{bmatrix} 0 \\ 0 \\ 0 \end{bmatrix},
\label{eq:F}
\end{equation}
where the first component of $F$ is to be interpreted as a relation
holding for all $\theta \in [0,2\pi]$ {and $|\A(\rho)|$ is the
  area inside the perturbed contour}. Infinite-dimensional
(continuous) system \eqref{eq:F} can be converted to an algebraic form
amenable to numerical solution with Newton's method by discretizing
the parameter $\theta$ uniformly as $\theta_j = (j-1)\Delta\theta$,
$j=1,\dots,M$, where $\Delta\theta = 2\pi /M$ and $M$ is an even
integer. The perturbation $\rho(\theta)$ is approximated in terms of a
truncated Fourier series
\begin{equation}
\rho(\theta) \approx \rho_N(\theta) := \sum_{k=0}^{N} a_k \cos(k\theta) + \sum_{k=1}^{N} b_k \sin(k\theta), 
\label{eq:rN}
\end{equation}
where $N := M / 2$. The cosine- and sine-series coefficients of the
shape perturbation $\rho_N$ can then be combined into a vector
\begin{equation}
\z = \left[a_0, a_1, \dots, a_N, b_1, \dots, b_N \right]^T \in \RR^{M+1}.
\label{eq:y}
\end{equation}
After collocating the first component of \eqref{eq:F},
cf.~\eqref{eq:CDa}, at the discrete parameter values $\theta_j$,
$j=1,\dots,M$, and using suitable discretizations (to be described
below), the resulting algebraic system takes the form
\begin{equation}
\bF\left(\begin{bmatrix} \z \\ W \end{bmatrix}\right)
= \begin{bmatrix} \mathbf{0} \\ 0 \\ 0 \end{bmatrix},
\label{eq:FF}
\end{equation}
where $\bF \; : \; \RR^{M+2} \rightarrow \RR^{M+2}$ and $\mathbf{0}
\in \RR^{M}$ is a column vector of zeros. Relation \eqref{eq:FF}
represents a system of nonlinear algebraic equations which can be
solved using Newton's method. Denoting $\bDelta\z$ and $\Delta W$ the
components of Newton's step corresponding to the updates,
respectively, of the series coefficients \eqref{eq:y} of the shape
perturbation and of the translation velocity $W$, a single iteration
of Newton's method requires solution of the system
\begin{subequations}
\label{eq:Newton}
\begin{align}
& \bnabla\bF\left(\begin{bmatrix} \z \\ W \end{bmatrix}\right)
\begin{bmatrix} \bDelta\z \\ \Delta W \end{bmatrix}
= -\bF\left(\begin{bmatrix} \z \\ W \end{bmatrix}\right), \label{eq:NewtonStep} \\
& \text{where} \quad \bnabla\bF\left(\begin{bmatrix} \z \\ W \end{bmatrix}\right) 
:= \begin{bmatrix} 
\left(\bM  - W \diag\left(\e_z\cdot\t_{\x(\theta_1,\dots,\theta_M)}\right) \bXi\right) & \left(\e_z\cdot\n_{\x(\theta_1,\dots,\theta_M)}\right) \label{eq:DF} \\
-[0, 2, \dots, 0, 2][0,\dots,0] & 0 \\
\left[U_0, U_1, \dots, U_N,V_1, \dots, V_N\right] & 0
\end{bmatrix}
\end{align}
\end{subequations}
is the Jacobian of the function $\bF$, cf.~\eqref{eq:FF}, in which the
expressions in the first column represent the partial derivatives of
the discretized kinematic condition \eqref{eq:CDa}, the normalization
condition \eqref{eq:CDb} and the fixed-area condition \eqref{eq:CDc}
with respect to the expansion coefficients \eqref{eq:y} obtained using
methods of the shape calculus \citep{dz01a}, cf.~\S \ref{sec:stab},
whereas the expressions in the second column are the {partial}
derivatives of these conditions with respect to the translation
velocity $W$. The vectors in the first row are to be understood as
$\e_z\cdot\t_{\x(\theta_1,\dots,\theta_M)} :=
[t_z(\theta_1),\dots,t_z(\theta_M)]^T$ and
$\e_z\cdot\n_{\x(\theta_1,\dots,\theta_M)} :=
[n_z(\theta_1),\dots,n_z(\theta_M)]^T$. Since the matrix $\bM$ also
arises in the discretization of the stability problem, evaluation of
its entries is discussed in detail in \S \ref{sec:numer},
cf.~\eqref{eq:M}. The term $W
\diag\left(\e_z\cdot\t_{\x(\theta_1,\dots,\theta_M)}\right) \bXi$,
where the matrix $\bXi$ is defined in \eqref{eq:Xi}, results from
shape differentiation of the second term on the left-hand side (LHS)
in condition \eqref{eq:CDa}. Noting that the shape differential of the
vortex area {$|\A(\rho)|$} is $|\A(\rho)|' = \left(\int_{\A}\,
  d\A \right)' = \int_{\partial\A} \rho \, ds$ \citep{dz01a}, the
partial derivatives of the fixed-area condition \eqref{eq:CDc} with
respect to expansion coefficients \eqref{eq:y} are obtained as $U_k :=
\int_0^{2\pi} \cos(k\theta') \frac{ds}{d\theta}(\theta')\, d\theta'$,
$V_k := \int_0^{2\pi} \sin(k\theta') \frac{ds}{d\theta}(\theta')\,
d\theta'$, $k=0,\dots,M$, where $s(\theta)$ is the arclength of the
contour.  Details concerning a spectrally-accurate approach to
evaluate such expressions are provided in \S \ref{sec:numer}.

The Jacobian $\bnabla\bF$ in \eqref{eq:DF} is singular with a
one-dimensional kernel space corresponding to arbitrary shifts of the
contour $\partial\A$ in the $z$ direction. In the process of solving
\eqref{eq:NewtonStep} to compute Newton's step {the
  singular-value decomposition (SVD) is used to ensure that the
  obtained Newton's step is orthogonal to this subspace.}

We emphasize that, as will be evident in \S \ref{sec:stab}, the
Jacobian $\bnabla\bF$ also contains key information necessary to set
up the generalized eigenvalue problem for the study of the stability
of vortex rings with respect to axisymmetric perturbations. This fact
constitutes the main practical advantage of the approach developed
here with respect to the method originally devised by
\citet{norbury-1973-JFM}.  Since the vortex rings are known to possess
a fore-and-aft symmetry, in ansatz \eqref{eq:rN} only the
cosine-series coefficients will be nonzero, i.e., $b_1 = \dots = b_N =
0$, {leading, via relation \eqref{eq:xeps}, to functions
  $r(\theta)$ and $z(\theta)$ which are, respectively, even and odd}.
However, we retain the sine series in subsequent derivations, because
their outcomes will be also used in the stability analysis in \S
\ref{sec:stab} where general perturbations without the fore-and-aft
symmetry will be considered.

Since for all $\alpha \in (0,\sqrt{2})$ the contours $\partial\A$ are
smooth \citep{norbury-1973-JFM}, we can expect spectral convergence of
the expansion \eqref{eq:rN}, i.e., $|a_k| \sim e^{-\beta k}$ for $k
\rightarrow \infty$ and some $\beta > 0$ \citep{n13}. However, since
the kinematic condition \eqref{eq:CDa} involves a convolution integral
which is discretized using a collocation approach, one may expect that
aliasing errors will arise \citep{b01}. They will manifest themselves
through an increase of $|a_k|$ with $k$ {when} $k > k_a$ for a
certain $k_a>0$. This issue is efficiently remedied by using an
implicit filtering approach in which Newton's step
\eqref{eq:NewtonStep} is redefined as
\begin{equation}
\bnabla\bF\left(\begin{bmatrix} \z \\ W \end{bmatrix}\right)
\begin{bmatrix} \bG^{\delta} \bDelta\z \\ \Delta W \end{bmatrix}
= -\bF\left(\begin{bmatrix} \z \\ W \end{bmatrix}\right)
\label{eq:NewtonStep2} 
\end{equation}
in which $\bG^{\delta} \in \RR^{(M+1)\times(M+1)}$ is a smoothing
operator given by a low-pass filter, here represented by the following
diagonal matrix
\begin{equation}
\bG^{\delta}  := \diag\left(0,
\left[\frac{1}{1+(\delta k)^{2p}}, \ k=1,\dots,M\right],
\left[\frac{1}{1+(\delta k)^{2p}}, \ k=1,\dots,M\right]\right).
\label{eq:Gd}
\end{equation}
In computations we will take $p=4$ as the order of the filter, whereas
the cutoff length-scale $\delta$ will be refined with the resolution
$M$ as $\delta = 2 \pi \gamma / M$ for some $\gamma > 1$
{ensuring that filtering does not affect the spectral accuracy of
  the approach.}

As the vortex rings become ``fatter'' in the limit $\alpha \rightarrow
\sqrt{2}$, the shape of their boundary $\partial A$ deviates from the
circle and develops ``corners'' characterized by an increasing
curvature. In order to resolve these features, increased resolutions
must be used varying from $M=32$ when $\alpha = 0.1$ to $M=4096$ when
$\alpha = 1.4$. {We have carefully checked the convergence of the
  computed shapes of the vortex boundary $\partial \A$ and of the
  translation velocity $W$ with refinement of the resolution $M$.}
The iterations of Newton's method are initiated with Norbury's
original solutions documented in \citet{norbury-1973-JFM}, or with
neighboring solutions for values of $\alpha$ not considered by
\citet{norbury-1973-JFM}. The iterations are declared converged when
the following stringent termination conditions are simultaneously
satisfied
\begin{equation}
\Bigg|\Bigg| \bF\left(\begin{bmatrix} \z \\ W \end{bmatrix}\right) \Bigg|\Bigg|_2 < \widetilde{\varepsilon}
\quad \text{and} \quad 
\Bigg|\Bigg| \bF\left(\begin{bmatrix} \z \\ W \end{bmatrix}\right) \Bigg|\Bigg|_{\infty} < \widetilde{\varepsilon},
\label{eq:NewtonTerm} 
\end{equation}
where the tolerance was set as $\widetilde{\varepsilon} = 10^{-10}$
for $\alpha \in (0,1.2]$ and was gradually relaxed to
$\widetilde{\varepsilon} = 10^{-5}$ as $\alpha$ increased to 1.4. The
reason for relaxing {the} tolerance is that at higher resolutions
required to represent the shapes of the vortex boundary when $\alpha >
1.2$, the accuracy of the solution is ultimately limited by the poor
algebraic conditioning of the (desingularized) Jacobian matrix
\eqref{eq:DF}. More specifically, the value of
$\widetilde{\varepsilon}$ roughly reflects the accuracy with which
Newton's step $\left[ \bG^{\delta}\bDelta\z^T,\Delta W \right]^T$ can
be evaluated {by solving problem \eqref{eq:NewtonStep2}} at a
given resolution.  In all cases convergence was typically achieved in
$\mathcal{O}(10)$ iterations, although due to the singularity of the
Jacobian \eqref{eq:DF} convergence was only linear. In the cases when
Norbury's original solutions were used as the initial guesses, the
residuals \eqref{eq:NewtonTerm} would typically {decrease} by
8--10 orders of magnitude.

\begin{figure}
\centering
\mbox{\hspace*{-0.85cm}
\subfigure[]{\includegraphics[width=0.66\textwidth]{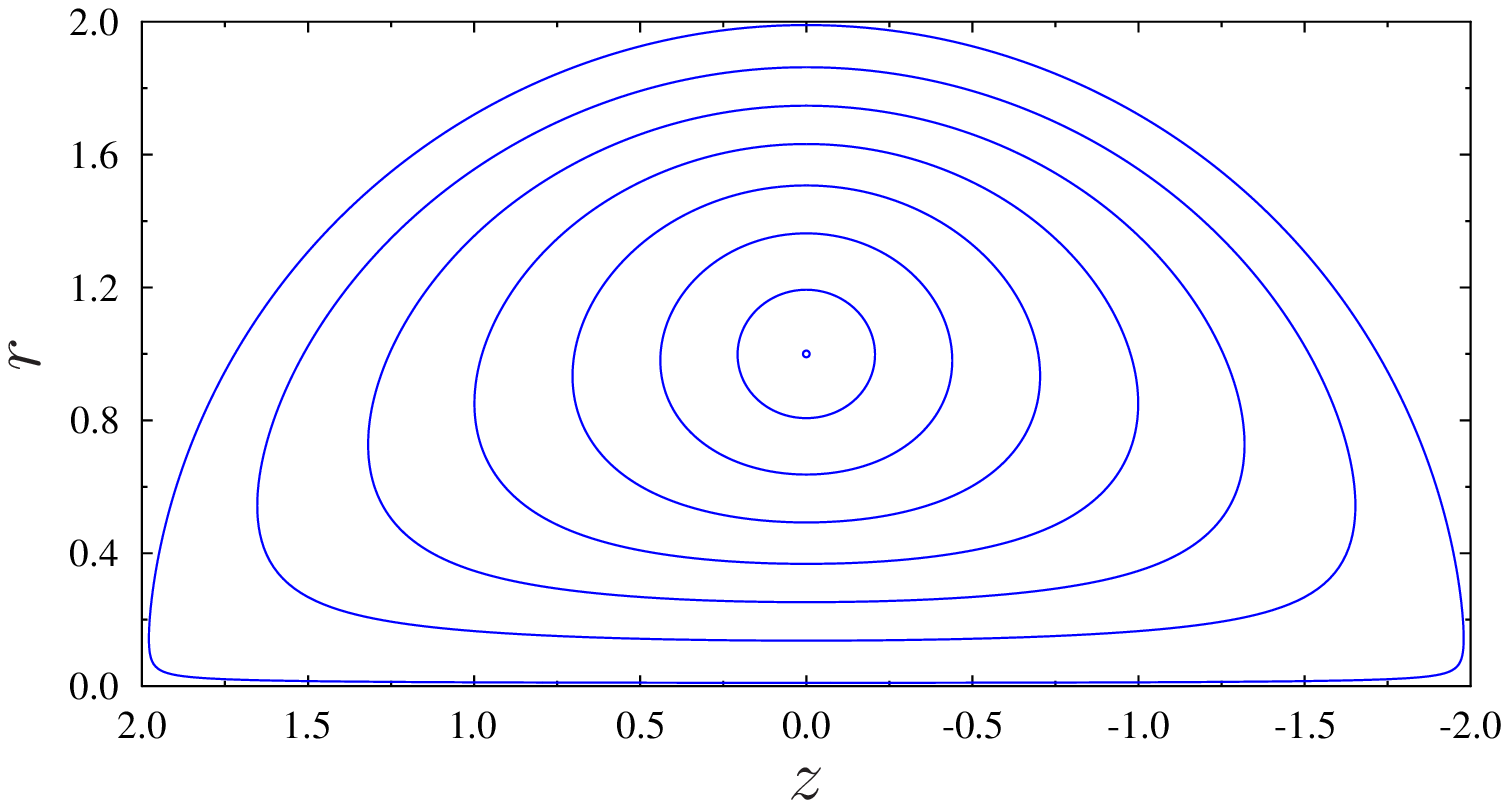}}
\hspace*{-0.75cm}
\subfigure[]{\includegraphics[width=0.5\textwidth]{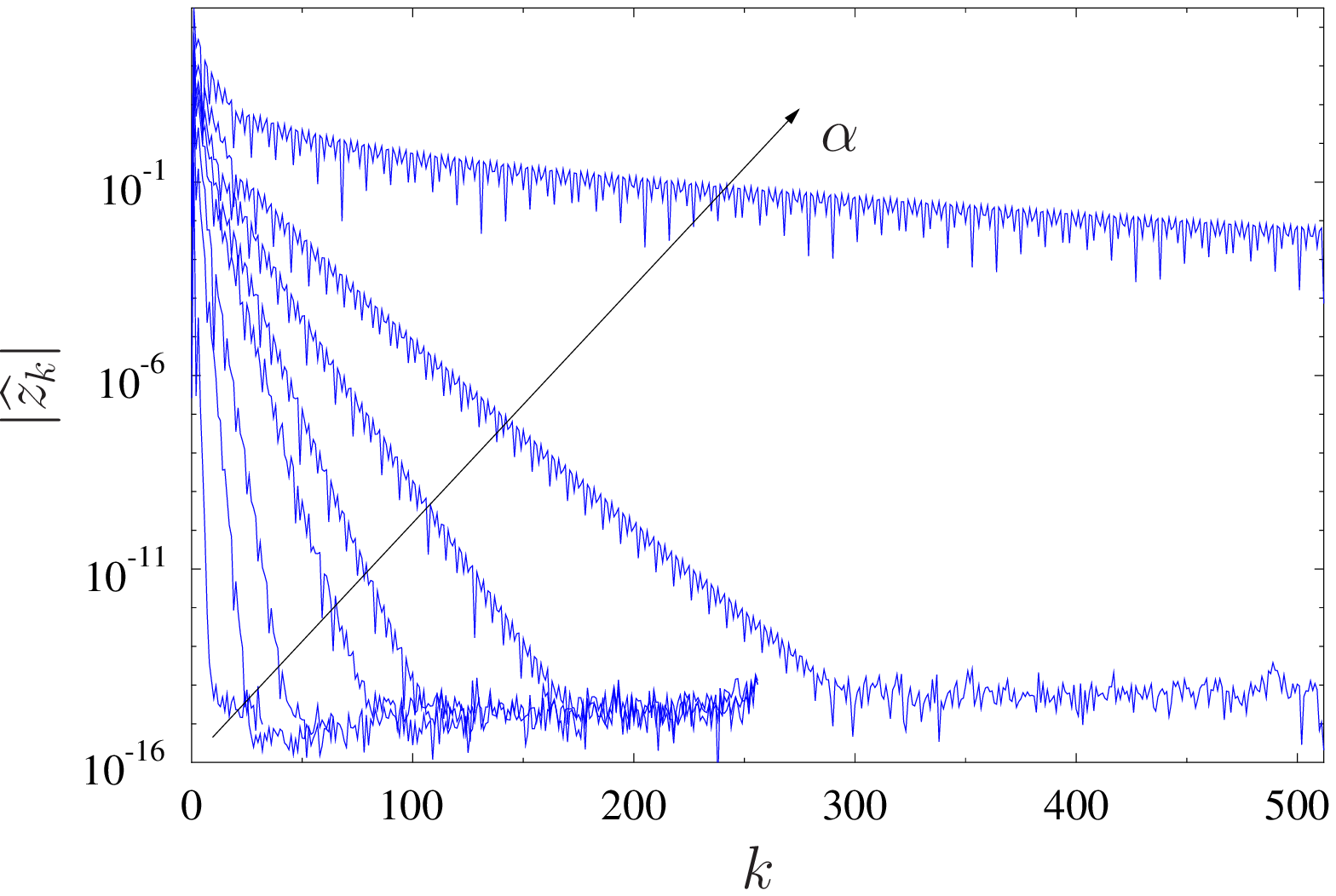}}}
\caption{(a) Vortex boundaries $\partial\A$ obtained by solving system
  \eqref{eq:FF} for $\alpha = 0.01, 0.2, 0.4, 0.6, 0.8, 1.0, 1.2,
  1.4$. (b) The amplitudes of the Fourier coefficients
  $\widehat{x}_k$, $k=1,\dots,N$, of the corresponding
  parametrizations $r(\theta)$ obtained for the same values of
  $\alpha$ (the trend with the increase of $\alpha$ is indicated with
  an arrow). {The resolutions used in the computations varied
    from $N=16$ for small $\alpha$ to $N=2048$ for $\alpha \rightarrow
    \sqrt{2}$.}}
\label{fig:vortices}
\end{figure}
The shapes of the vortex boundary $\partial\A$ obtained as solutions
to \eqref{eq:FF} for representative values of $\alpha$ are presented
in figure \ref{fig:vortices}(a) where a {gradual} evolution of
the shape from a thin vortex filament to Hill's ``fat'' vortex is
evident as $\alpha$ increases from 0.01 to nearly $\sqrt{2}$. {To
  simplify notation, hereafter the equilibrium shapes of the vortex
  boundary $\partial \A$ are denoted $\x = [r,z]^T$, with the
  coordinates given in terms of parametrizations $r = r(\theta)$ and
  $z = z(\theta)$.}  The amplitudes of the {corresponding}
Fourier coefficients $\widehat{z}_k$, $k=1,\dots,N$, are shown in
figure \ref{fig:vortices}(b) (the spectra of $r(\theta)$ are
essentially indistinguishable and are not shown). We see that for all
values of $\alpha$ the amplitudes $|\widehat{z}_k|$ of the Fourier
coefficients decay exponentially with the wavenumber $k$, although
this property is gradually lost as $\alpha \rightarrow \sqrt{2}$ where
the vortex boundary $\partial\A$ develops a sharp corner (such that
$r(\theta)$ and $z(\theta)$ are no longer smooth functions of
$\theta$).  {In figure \ref{fig:vortices}(b) we also observe
  that, interestingly, the dependence of $|\widehat{z}_k|$ on the
  wavenumber $k$ is characterized by a certain recurrent pattern.}
The translation velocity $W$ is positive for all values of $\alpha$,
meaning that the vortices shown in figure \ref{fig:vortices}(a) move
to the {left}. When $\alpha \rightarrow 0$, this translation
velocity becomes unbounded as $(1/4)(\ln(8 / \alpha) - 1/4)$, whereas
when $\alpha \rightarrow \sqrt{2}$, $W$ approaches $4/15$, both of
which are well-known results for the propagation velocity of an
infinitely thin vortex ring and Hill's vortex, respectively
\citep{wu-book-2006}.  As regards the integral diagnostics, the
variation of the vortex area $|\A|$, its circulation $\Gamma$ and its
{hydrodynamic} impulse $\I$ with $\alpha$ is consistent with the
behavior of these quantities discussed in \citet{norbury-1973-JFM},
both as $\alpha \rightarrow 0$ and $\alpha \rightarrow \sqrt{2}$. The
data $\{r_j,z_j\}$, where $r_j = r(\theta_j)$ and $z_j = z(\theta_j)$,
$j=1,\dots,M$, describing the vortex boundaries $\partial\A$ for a
range of different $\alpha \in (0,\sqrt{2})$ is provided in
Supplementary Information together with the corresponding values of
the propagation velocity $W$.  The linear stability of these
equilibrium configurations is discussed in the next section.

\section{Linear Stability Analysis}
\label{sec:stab}

In this section we derive a singular integro-differential equation
characterizing the linear response of the boundary $\partial\A(t)$ of
Norbury's vortices, cf.~figure \ref{fig:vortices}(a), to axisymmetric
perturbations.  This will be done by applying the tools of the
shape-differential calculus \citep{dz01a} to the contour-dynamics
formulation \eqref{eq:v} of the problem. Finally, we obtain a
constrained eigenvalue problem from which the stability properties of
Norbury's vortices can be deduced.

In addition to the circulation, impulse and energy conserved by all
classical solutions of Euler's equations, axisymmetric inviscid flows
also conserve {Casimirs $\iint_{\A(t)} \Psi({\omega} / {r}) \,r \,
  d\A$, where $\Psi \; : \; \RR \rightarrow \RR$ is an arbitrary
  function with sufficient regularity \citep{mohseni-2001-PF}.}  Since
circulation is {a particularly simple Casimir, defined by
  $\Psi(\xi) = \xi$ for some $\xi \in \RR$,} we will focus on
stability analysis with respect to perturbations preserving this
quantity, which is the same approach as was also taken by
\citet{mm78,ProtasElcrat2016} in their studies of the stability of
Hill's vortex.  {By this we mean that the perturbed vortex ring
  will have the same circulation as the corresponding vortex ring in
  the equilibrium configuration (however, under the linearized Euler
  evolution, this circulation will in general change).  The}
circulation $\Gamma$ of the flow in the meridional plane is defined as
\begin{equation}
\Gamma := \iint_{\A(t)} {\omega} \, d\A
\label{eq:circ}
\end{equation}
and, since the flows considered here have the property $\omega = \C
r$, cf.~\eqref{eq:Euler3D}--\eqref{eq:f}, conservation of circulation
\eqref{eq:circ} implies the conservation of the volume of the vortical
region. {We add that restricting the vorticity of the perturbed
  vortex to the form $\omega = \C r$ is justified by recognizing that
  such configurations can be obtained as rearrangements of the
  vorticity distribution in the reference configuration
  \citep{Benjamin1975}.}

{We now describe how axisymmetric perturbations are applied to
  the equilibrium configurations discussed in \S \ref{sec:rings},
  cf.~figure \ref{fig:vortices}(a).}  Following the ideas laid out by
\citet{ep13}, we now assume that the position of the vortex boundary
$\partial\A(t)$ depends on time $t$ and, to simplify calculations,
introduce another parametrization $\x = \x(t,s) \in \partial\A(t)$ for
it, now in terms of the arclength $s$ of the contour. Since the maps
$r = r(\theta)$ and $z = z(\theta)$ are assumed invertible and $ds =
\sqrt{(dr)^2+(dx)^2}$, we can rewrite ansatz \eqref{eq:xeps} as
\begin{equation}
\x^{\epsilon}(t,s) = \x(s) + \epsilon\, \rho(t,s) \, \n_{\x(s)},
\label{eq:xeps2}
\end{equation}
where $s = s(\theta)$ and the point $\x$ belongs to the boundary
$\partial\A$ of the equilibrium configuration. {We remark that
  for smooth reference contours $\x$ ansatz \eqref{eq:xeps2} does not
  restrict the class of admissible contour deformations except for the
  assumption that the perturbed contour $\x^{\epsilon}$ be ``close''
  to the reference contour $\x$ \citep{dz01a}, which is however
  consistent with how the linear stability problem is formulated.}
Using equation \eqref{eq:v} we then deduce
\begin{equation}
\n_{\x^{\epsilon}}\cdot\frac{d\x^{\epsilon}(t)}{dt} = 
{\n_{\x^{\epsilon}}\cdot}\v^{\epsilon}(\x^{\epsilon}(t)) = 
 \n_{\x^{\epsilon}}\cdot\left[ \C \,
\int_{\partial\A^{\epsilon}(t)} \bK(\x^{\epsilon}(t),\x')\, ds_{\x'} - W\, \e_z\right], 
\label{eq:vne}
\end{equation}
from which the equilibrium condition \eqref{eq:vn} is obtained by
setting $\epsilon = 0$. The equation describing the evolution of the
perturbation $\rho$ is obtained by linearizing relation \eqref{eq:vne}
around the equilibrium configuration characterized by \eqref{eq:vn}
which is equivalent to expanding \eqref{eq:vne} in powers of
$\epsilon$ and retaining the first-order terms. Since equation
\eqref{eq:vne} involves perturbed quantities defined on the perturbed
vortex boundary $\partial A^{\epsilon}(t)$, the proper way to obtain
this linearization is using methods of the shape-differential calculus
\citep{dz01a}. Below we state the main results only and refer the
reader to our earlier study \citep{ep13} for details of all
intermediate transformations.  Shape-differentiating the left-hand
side of relation \eqref{eq:vne} and setting $\epsilon = 0$ we obtain
\begin{equation}
\frac{d}{d\epsilon} \left[\n_{\x^{\epsilon}}\cdot\frac{d\x^{\epsilon}(t)}{dt}\right]\bigg|_{\epsilon=0} = \Dpartial{\rho}{t}.
\label{eq:vne_lhs}
\end{equation}
As regards the right-hand side (RHS) in \eqref{eq:vne}, we obtain
\begin{multline}
\frac{d}{d\epsilon} \left\{  \n_{\x^{\epsilon}}\cdot\left[\C \,
\int_{\partial\A^{\epsilon}(t)} \bK(\x^{\epsilon}(t),\x')\, ds_{\x'} - W\, \e_z\right] \right\}\Bigg|_{\epsilon=0} = \\
= - \Dpartial{\rho}{s} \, \v(\x)\cdot\t_{\x} 
+\C \, \rho(s)\, \n_{\x}\cdot\int_{\partial\A} \Dpartial{\bK}{n_{\x}}\, ds' 
+\C \, \n_{\x}\cdot \int_{\partial\A} \bK' + \left[\Dpartial{\bK}{n_{\x'}} + \varkappa_{\x'}\bK\right] \rho(s')\, ds',
\label{eq:vne_rhs}
\end{multline}
where $\varkappa_{\x}$ is the curvature of the contour $\partial\A$
and $\bK'$ is the shape differential of the kernel $\bK$,
cf.~\eqref{eq:K}. The orientation of the unit vectors $\t_{\x}$ and
$\n_{\x}$, and the sign of the curvature $\varkappa_{\x}$ satisfy
Frenet's convention. The terms in the integrand expressions in
\eqref{eq:vne_rhs} involving the derivatives of the kernel in the
directions of the normal vectors are expressed as $\partial \bK /
\partial n_{\x} = \n_{\x}\cdot\bnabla_{\x}\bK(\x,\x')$ and $\partial
\bK / \partial n_{\x'} = \n_{\x'}\cdot\bnabla_{\x'}\bK(\x,\x')$, where
the subscript on the operator $\bnabla$ indicates which variable
differentiation is performed with respect to, whereas the shape
differential of the kernel is evaluated as follows
\begin{equation}
\begin{aligned}
\bK'(r,z,r',z') 
& = - r'  H(r,z,r',z') \, \frac{d}{d\epsilon} n^{\epsilon}_z(r',z')\big|_{\epsilon=0} \, \e_{r}  \\
& \phantom{=} -\left[ (z'-z) \, G(r,z,r',z') \, \frac{d}{d\epsilon} n^{\epsilon}_z(r',z')\big|_{\epsilon=0} - 
r \, H(r,z,r',z') \,  \frac{d}{d\epsilon} n^{\epsilon}_{r}(r',z')\big|_{\epsilon=0} \right]\, \e_z \\
& = r'  H(r,z,r',z') \,  n_{r}(r',z') \,\Dpartial{\rho}{s} \, \e_{r}  \\
& \phantom{=} + \left[ (z'-z) \, G(r,z,r',z') \, n_{r}(r',z') + 
r \, H(r,z,r',z') \,  n_{x}(r',z') \right] \, \Dpartial{\rho}{s} \, \e_z \\
& =: \DK'(r,z,r',z') \, \Dpartial{\rho}{s},
\end{aligned}
\label{eq:dK}
\end{equation}
where the second equality follows from the expression for the shape
derivative of the unit normal vector, i.e., $d\n^{\epsilon} /
d\epsilon \,|_{\epsilon=0} = - (\partial \rho / \partial s) \, \t$, and
Frenet's identities \citep{ep13}.

As explained by \citet{ep13}, the three terms on the RHS of
\eqref{eq:vne_rhs} represent the shape-sensitivity of the RHS of
\eqref{eq:vne} to perturbations \eqref{eq:xeps2} applied separately to
the normal vector $\n_{\x}$, the evaluation point $\x$ and the contour
$\partial\A$ over which the integral is defined. Since the flow
evolution is subject to constraint \eqref{eq:circ}, this will restrict
the admissible perturbations $\rho$. Indeed, noting that $\omega = \C
r$, cf.~\eqref{eq:Euler3D}--\eqref{eq:f}, and shape-differentiating
relation \eqref{eq:circ} we obtain the following condition
\citep{ep13}
\begin{equation}
\int_{\partial\A} {{r} \rho(0,s')}\, ds' = 0
\label{eq:r0}
\end{equation}
restricting the class of admissible perturbations to those which do
not modify the circulation \eqref{eq:circ} (analogous conditions {can}
be obtained when the perturbations are constructed to preserve other
integral invariants mentioned above). We note that the constraint
imposed in the formulation of the stability problem,
cf.~\eqref{eq:circ}, is different from the constraints which were
imposed during the computation of the equilibrium configurations,
cf.~\eqref{eq:CDb}--\eqref{eq:CDc}, which {resulted} from the
assumed normalization.

It is now convenient to return to the original parametrization in
terms of $\theta$, i.e., assume again that $r = r(\theta)$, $z =
z(\theta)$, $r' = r'(\theta')$ and $z' = z'(\theta')$,
such that combining relations \eqref{eq:vne_lhs}--\eqref{eq:r0} and
replacing the line integrals with the corresponding definite ones we
finally obtain the equation describing the deformation $\rho(t,\theta)$
of the vortex boundary $\partial\A(t)$, cf.~\eqref{eq:xeps2}, in
response to axisymmetric perturbations in the linear regime
\begin{subequations}
\label{eq:L}
\begin{align}
 \Dpartial{\rho}{t} & = 
- \C \, \Dpartial{\rho}{\theta} \left(\frac{ds}{d\theta}\right)^{-1} \, \int_0^{2\pi} I_0(\theta,\theta') \frac{ds}{d\theta}(\theta')\, d\theta'
+\C \, \rho(\theta)\, \int_0^{2\pi} I_1(\theta,\theta')\frac{ds}{d\theta}(\theta')\, d\theta' \nonumber \\  
& \phantom{=} +\C \, \int_0^{2\pi} I_2(\theta,\theta')\rho(\theta') \frac{ds}{d\theta}(\theta')\, d\theta' 
+ \C \, \int_0^{2\pi} I_3(\theta,\theta') \Dpartial{\rho}{\theta}(\theta')\, d\theta'
\label{eq:La} \\
& =: \left({ \G} \rho \right)(\theta) \nonumber \\
& \textrm{subject to:} \quad \int_0^{2\pi} r(\theta') \rho(\theta') \frac{ds}{d\theta}(\theta')\, d\theta' = 0,
\label{eq:Lb}
\end{align}
\end{subequations}
where {$\G$} denotes the associated linear operator and 
\begin{subequations}
\label{eq:I}
\begin{align}
I_0(\theta,\theta') & := \t_\theta \cdot \bK(\theta,\theta'), \label{eq:I0} \\
I_1(\theta,\theta') & := \n_\theta \cdot \Dpartial{\bK(\theta,\theta')}{n_{\theta}},  \label{eq:I1} \\
I_2(\theta,\theta') & := \n_\theta \cdot \left[\Dpartial{\bK(\theta,\theta')}{n_{\theta'}} + \varkappa_{\theta'}\bK(\theta,\theta')\right], \label{eq:I2} \\
I_3(\theta,\theta') & := \n_\theta \cdot \DK(\theta,\theta'). \label{eq:I3}
\end{align}
\end{subequations}
Given the form of the kernel $\bK$, cf.~\eqref{eq:K}, expressions
\eqref{eq:I0}--\eqref{eq:I3} take rather complicated forms as
functions of $r(\theta)$, $z(\theta)$, $r'(\theta')$, and
$z'(\theta')$. They can be however readily obtained using symbolic
algebra tools such as {\tt Maple}.

As regards the singularities of the kernels, one can verify by
inspection that
\begin{subequations}
\label{eq:Ksing}
\begin{align}
\forall \ \theta, \theta'  \in [0,2\pi] \quad
& \lim_{\theta' \rightarrow \theta} \, \n_\theta \cdot \bK(\theta,\theta') = 0, \label{eq:Ksingn} \\
& \lim_{\theta' \rightarrow \theta} \, \frac{I_0(\theta,\theta')}{\ln\sin^2\left(\frac{\theta-\theta'}{2}\right)} = 
- \frac{1}{4\pi} r(\theta), \label{eq:Ksing0} \\
& \lim_{\theta' \rightarrow \theta} \, \frac{I_1(\theta,\theta')}{\ln\sin^2\left(\frac{\theta-\theta'}{2}\right)} = 
- \frac{1}{4\pi} n_z(\theta), \label{eq:Ksing1} \\
& \lim_{\theta' \rightarrow \theta} \, \frac{I_2(\theta,\theta')}{\ln\sin^2\left(\frac{\theta-\theta'}{2}\right)} = 
  \frac{1}{4\pi} n_z(\theta), \label{eq:Ksing2} \\
& \lim_{\theta' \rightarrow \theta} \, \frac{I_3(\theta,\theta')}{\ln\sin^2\left(\frac{\theta-\theta'}{2}\right)} = 
- \frac{1}{4\pi} r(\theta), \label{eq:Ksing3}
\end{align}
\end{subequations}
where the expression in the denominators on the LHS is chosen to
combine the correct singular behavior for $\theta' \rightarrow \theta$
with $2\pi$-periodic dependence on $\theta$ and $\theta'$.
{Relations \eqref{eq:Ksing0}--\eqref{eq:Ksing3} are deduced using
  known properties of the complete elliptic integrals appearing in
  \eqref{eq:K} \citep{olbc10}, and then computing the limits
  symbolically using the software package {\tt Maple}.}  From
\eqref{eq:Ksingn} we conclude that the singularity of the normal
component of the kernel $\bK$ is removable, whereas all the integrals
in \eqref{eq:La} are to be understood in the improper sense.
Properties \eqref{eq:Ksing0}--\eqref{eq:Ksing3} will be instrumental
in achieving spectral accuracy in the discretization of system
\eqref{eq:L}.

After introducing the ansatz 
\begin{equation}
{\rho(t,\theta) = e^{i \lambda t}\,u(\theta) + C.C.,} 
\label{eq:r}
\end{equation}
where $i:=\sqrt{-1}$ and $\lambda \in \CC$, system \eqref{eq:L} takes
the form of a constrained eigenvalue problem
\begin{subequations}
\label{eq:evalp}
\begin{alignat}{2}
& i \, \lambda \, u(\theta) = \left( \G u\right)(\theta), & \quad & \theta \in [0,2\pi] \label{eq:evalp1} \\
& \textrm{subject to:} \quad \text{periodic boundary conditions}, & &  \label{eq:evalp2} \\
& \phantom{\textrm{subject to:}} \quad \int_0^{2\pi} r(\theta') \rho(\theta') \frac{ds}{d\theta}(\theta')\, d\theta' = 0, & & \label{eq:evalp3}
\end{alignat}
\end{subequations}
where the operator $\G$ is defined in \eqref{eq:La}. The eigenvalues
$\lambda$ and the corresponding eigenvectors $u$ characterize the
stability of Norbury's vortex rings, cf.~figure \ref{fig:vortices}(a),
to axisymmetric perturbations. Our approach to the numerical solution
of the constrained eigenvalue problem \eqref{eq:evalp} is presented in
the next section.

\section{Numerical Approach}
\label{sec:numer}

In this section we describe our numerical approach focusing on the
discretization of system \eqref{eq:evalp} and the solution of the
resulting algebraic eigenvalue problem. We will also provide some
details about how these techniques have been validated. Our approach
is an adaptation to the present problem of the methodology validated
and thoroughly tested in earlier studies
\citep{ep13,ProtasElcrat2016}. It also shares a number of key
ingredients with the methodology employed to approximate Newton's step
\eqref{eq:NewtonStep} in the solution of problem \eqref{eq:F} when
computing equilibrium configurations in \S \ref{sec:rings}.

As a starting point, we recall that the equilibrium configurations of
the vortex boundary $\partial\A$ are represented in terms of the
parametrization $[r(\theta),z(\theta)]$, $\theta\in[0,2\pi]$,
discretized (with spectral accuracy) as $\{ r_j,z_j\}$, $j=1,\dots,M$.
The eigenvectors $u$ are then approximated with a truncated series
similar to the one already used in \eqref{eq:rN}, i.e.,
\begin{equation}
u(\theta) \approx u_N(\theta) := \sum_{k=0}^{N} a_k \cos(k\theta) + \sum_{k=1}^{N-1} b_k \sin(k\theta)
\label{eq:uN}
\end{equation}
(without the risk of confusion, the expansion coefficients in
\eqref{eq:rN} and \eqref{eq:uN} are denoted with the same symbols).
{Since the total number of coefficients parameterizing
  $u_N(\theta)$ in \eqref{eq:uN} must be equal to $M = 2N$, we
  arbitrarily choose to truncate the sine series in \eqref{eq:uN} at
  $k=N-1$.}  After substitution of ansatz \eqref{eq:uN}, equation
\eqref{eq:evalp1} is collocated on the uniform grid
$\theta_1,\dots\theta_M$, whereas constraint \eqref{eq:evalp3} takes
the form
\begin{equation}
\begin{aligned}
a_0 \, \int_0^{2\pi} r(\theta') \frac{ds}{d\theta}(\theta')\, d\theta'
& + \sum_{k=1}^N a_k \, \int_0^{2\pi} r(\theta') \cos(k\theta')\frac{ds}{d\theta}(\theta')\, d\theta' \\
& + \sum_{k=1}^{N-1} b_k \, \int_0^{2\pi} r(\theta') \sin(k\theta')\frac{ds}{d\theta}(\theta')\, d\theta' = 0.
\end{aligned}
\label{eq:c}
\end{equation}
The factor $ds / d \theta = \sqrt{\left(dr(\theta)/d\theta \right)^2 +
  \left(dz(\theta)/d\theta \right)^2 }$ can be approximated at
the collocation points $\theta_j$, $j=1,\dots,M$, as $\widetilde{ds /
  d \theta}|_j$ with spectral accuracy applying a suitable
differentiation matrix for periodic functions
\citep{trefethen:SpecMthd} to the discrete representations of the
parametrizations $r(\theta)$ and $z(\theta)$. Approximating the
integrals in \eqref{eq:c} using the trapezoidal quadrature, which for
smooth integrand expressions defined on periodic domains achieves
spectral accuracy, we obtain the constraint vector $\c =
[c_1,\dots,c_M]$, where
\begin{equation}
\begin{aligned}
c_1 := \Delta\theta \sum_{j=1}^M r_j \widetilde{\frac{ds}{d\theta}}\Big|_j, \quad
c_{1+k} & := \Delta\theta \sum_{j=1}^M r_j \cos(k\theta'_j) \widetilde{\frac{ds}{d\theta}}\Big|_j, \quad 
k = 1, \dots,N, \\
c_{N+1+k} & := \Delta\theta \sum_{j=1}^M r_j \sin(k\theta'_j) \widetilde{\frac{ds}{d\theta}}\Big|_j, \quad
k = 1, \dots,N-1.
\end{aligned}
\label{eq:c2}
\end{equation}
We also introduce the notation $\w = [w_1,\dots,w_{M}]^T$, where
$w_1 := a_0$, $w_{1+k} := a_k$, $k=1,\dots,N$, and $w_{N+1+k} := b_k$, $k = 1,
\dots,N-1$.  As a result, we obtain the following discretization of the  generalized
eigenvalue problem \eqref{eq:evalp}
\begin{subequations}
\label{eq:evald}
\begin{align}
i \lambda \sum_{k=1}^{M} A_{jk} w_k 
& = \sum_{k=1}^{M} \left( B_{jk} + C_{jk} + D_{jk} + E_{jk} \right) w_k,
\quad  \quad j=1,\dots,M, \label{eq:evalda} \\
\sum_{k=1}^{M} c_k w_k & = 0 \label{eq:evaldb}
\end{align}
\end{subequations}
in which 
\begin{equation}
A_{jk} := 
\begin{cases}
\cos \left[(k-1)\theta_j\right],  & 1 \le k \le N+1 \\
\sin \left[(k-N-1)\theta_j\right],  & N+1 < k \le M 
\end{cases},
\qquad j=1,\dots,M,
\label{eq:Ajk}
\end{equation}
is an (invertible) collocation matrix related to ansatz \eqref{eq:uN}
and the uniform discretization of $\theta$, whereas the matrices $\bB$,
$\bC$, $\bD$ and $\bE$ correspond to the four terms in operator $\G$,
cf.~\eqref{eq:La}. We note that, as stated above, problem
\eqref{eq:evald} appears overdetermined because there are $M+1$
conditions and only $M$ degrees of freedom $w_1,\dots,w_M$.  However,
this apparent inconsistency will be remedied further below in this
section.

The entries of matrix $\bB$, corresponding to the first term in
operator $\G$, are defined as follows (there is {\em no} summation
when indices are repeated)
\begin{align}
& B_{jk} := \C \, \left[ \widetilde{\frac{ds}{d\theta}}\Big|_j \right]^{-1} \, 
\left(\int_0^{2\pi} I_0(\theta_j,\theta') \frac{ds}{d\theta}(\theta')\, d\theta'\right) \, 
\Xi_{jk}, \quad j,k=1,\dots,M, \label{eq:Bjk} \\ 
& \text{where} \ \Xi_{jk} := 
\begin{cases}
- (k-1) \sin \left[(k-1)\theta_j\right],  & 1 \le k \le N+1 \\
(k-N-1) \cos \left[(k-N-1)\theta_j\right],  & N+1 < k \le M 
\end{cases},
\qquad j=1,\dots,M.
\label{eq:Xi}
\end{align}
The improper integral in \eqref{eq:Bjk}, representing the tangential
velocity component, cf.~\eqref{eq:I0}, can be evaluated using property
\eqref{eq:Ksing0} to separate the singular part of the kernel as
\citep{h95}
\begin{equation}
\begin{aligned}
\int_0^{2\pi} I_0(\theta,\theta') \frac{ds}{d\theta}(\theta')\, d\theta' =: 
& \underbrace{\int_0^{2\pi} \left[I_0(\theta,\theta') + \frac{r(\theta)}{4\pi} \ln\sin^2\left(\frac{\theta-\theta'}{2}\right)\right]\frac{ds}{d\theta}(\theta')\, d\theta'}_{\T_0(\theta)}  \\  
& - \frac{r(\theta)}{4\pi} \underbrace{ \int_0^{2\pi}\ln\sin^2\left(\frac{\theta-\theta'}{2}\right) \frac{ds}{d\theta}(\theta')\, d\theta'}_{\Phi_0(\theta)}.
\end{aligned}
\label{eq:intI0}
\end{equation}
Here $\T_0(\theta)$ has a bounded and continuous integrand 
and can be therefore approximated with spectral accuracy using the
trapezoidal quadrature noting also that for all $\theta \in [0,2\pi]$
\begin{equation}
\lim_{\theta' \rightarrow \theta} \left[I_0(\theta,\theta') + \frac{r(\theta)}{4\pi} \ln\sin^2\left(\frac{\theta-\theta'}{2}\right)\right]
= - \frac{r(\theta)}{4\pi}\left[\ln\left(\frac{\frac{ds}{d\theta}}{r}\right)^2 + 4(1 - \ln 2)\right].
\label{eq:intI00}
\end{equation}
{This and analogous relations below, cf.~\eqref{eq:intI10},
  \eqref{eq:intI20} and \eqref{eq:intI30}, are obtained symbolically
  using the software package {\tt Maple} by considering generalized
  series expansions of the expression on the LHS.}

In order to evaluate the factor $\Phi_0(\theta)$ in the singular part
of \eqref{eq:intI0}, we expand $ds/d\theta$ in a truncated
trigonometric series
\begin{equation}
\frac{ds}{d\theta}(\theta) = \sum_{k=0}^{N} \eta_k \cos(k\theta) + \zeta_k \sin(k\theta),
\label{eq:dsdtN}
\end{equation}
where the coefficients $\{\eta_k, \zeta_k\}_{k=0}^N$ can be evaluated
based on the discrete values $\widetilde{ds / d \theta}|_j$,
$j=1,\dots,M$, using FFT (while $\zeta_0 = \zeta_N \equiv 0$, for
compactness of notation, we will include these terms in different
expressions below).  Then,
\begin{equation}
\Phi_0(\theta) = \int_0^{2\pi}\ln\sin^2\left(\frac{\theta-\theta'}{2}\right) \frac{ds}{d\theta}(\theta')\, d\theta'
= \sum_{k=0}^{N} \eta_k C_k(\theta) + \zeta_k S_k(\theta),
\label{eq:phi0}
\end{equation}
where 
\begin{subequations}
\label{eq:CSk}
\begin{align}
C_k(\theta) &:= \int_0^{2\pi} \cos(k\theta') \ln\sin^2\left(\frac{\theta-\theta'}{2}\right) \, d\theta'
= \begin{cases}
-4\pi \ln 2 & k = 0 \\
-\frac{2 \pi}{k} \cos(k\theta) & k \ge 1
\end{cases},
\label{eq:Ck} \\
S_k(\theta) &:= \int_0^{2\pi} \sin(k\theta') \ln\sin^2\left(\frac{\theta-\theta'}{2}\right) \, d\theta'
= \begin{cases}
0  & k = 0 \\
-\frac{2 \pi}{k} \sin(k\theta) & k \ge 1
\end{cases}.
\label{eq:Sk}
\end{align}
\end{subequations}
The improper integrals defining $C_k(\theta)$ and $S_k(\theta)$ can be
evaluated using methods of complex analysis and the proofs of
identities \eqref{eq:Ck}--\eqref{eq:Sk} are provided in Appendix
\ref{sec:CSk}.  Relations \eqref{eq:intI0}--\eqref{eq:CSk} thus define
a spectrally-accurate approach to approximate the integral in
\eqref{eq:Bjk}.

The entries of matrix $\bC$, corresponding to the second term in
operator $\G$, are defined as follows
\begin{equation}
C_{jk} := \left(\C\, \int_0^{2\pi} I_1(\theta_j,\theta') \frac{ds}{d\theta}(\theta')\, d\theta'\right) \, A_{jk}, 
\qquad j,k = 1,\dots,M,
\label{eq:Cjk}
\end{equation}
where the coefficient is given by an improper integral evaluated as
above, i.e., using property \eqref{eq:Ksing1} to split the integral
into a regular and a singular part
\begin{equation}
\begin{aligned}
\int_0^{2\pi} I_1(\theta,\theta') \frac{ds}{d\theta}(\theta')\, d\theta' =: 
& \underbrace{\int_0^{2\pi} \left[I_1(\theta,\theta') + \frac{n_z(\theta)}{4\pi} \ln\sin^2\left(\frac{\theta-\theta'}{2}\right)\right]\frac{ds}{d\theta}(\theta')\, d\theta'}_{\T_1(\theta)}  \\  
& - \frac{n_z(\theta)}{4\pi} \underbrace{ \int_0^{2\pi}\ln\sin^2\left(\frac{\theta-\theta'}{2}\right) \frac{ds}{d\theta}(\theta')\, d\theta'}_{\Phi_0(\theta)}.
\end{aligned}
\label{eq:intI1}
\end{equation}
The regular part $\T_1(\theta)$ can be approximated with the
trapezoidal quadrature noting that
\begin{multline}
\lim_{\theta' \rightarrow \theta} \left[I_1(\theta,\theta') + \frac{n_z(\theta)}{4\pi} \ln\sin^2\left(\frac{\theta-\theta'}{2}\right)\right]
= \frac{-1}{4\pi \left(\frac{ds}{d\theta}\right)^2 }
\Bigg\{ \left[ 6 n_{r}^2 + \ln\left(\frac{\frac{ds}{d\theta}}{4r}\right)^2 \right] \left(\frac{ds}{d\theta}\right)^2 n_z \\
+ 2 \frac{d n_z}{d\theta} r n_{r}  \left( \frac{d z}{d\theta} n_z +  \frac{d r}{d\theta} n_{r}\right) 
+ 2 \frac{d n_{r}}{d\theta} r n_{r}  \left( \frac{d z}{d\theta} n_{r} -  \frac{d r}{d\theta} n_{z}\right) 
- 2 \frac{d n_{r}}{d\theta} \frac{d z}{d\theta} r \\
-2 \frac{d z}{d\theta} \left[ \frac{d z}{d\theta} n_z \left(2 n_{r}^2 - 1\right) + 
2 \frac{d r}{d\theta} n_{r} \left( n_{r}^2 -1 \right)\right]
\Bigg\} =: \Q(\theta),
\label{eq:intI10}
\end{multline}
whereas approximation of the singular part is already given by
relations \eqref{eq:phi0}--\eqref{eq:CSk}.

The entries of matrix $\bD$, corresponding to the third term in
operator $\G$, are defined as follows
\begin{equation}
D_{jk} := \C\, 
\begin{cases}
\int_0^{2\pi} I_2(\theta_j,\theta')\cos \left[(k-1)\theta'\right] \frac{ds}{d\theta}(\theta')\, d\theta', & 1 \le k \le N+1 \\
\int_0^{2\pi} I_2(\theta_j,\theta')\sin \left[(k-N-1)\theta'\right] \frac{ds}{d\theta}(\theta')\, d\theta', & N+1 < k \le M
\end{cases},
\quad j=1,\dots,M,
\label{eq:Djk}
\end{equation}
which represents the action of a weakly-singular integral operator on
trigonometric functions and can be evaluated as above using property
\eqref{eq:Ksing2} to split the integrals in \eqref{eq:Djk} into
regular and singular parts. For the integrals involving the cosine
functions  we thus obtain when $k \ge 0$
\begin{equation}
\begin{aligned}
\int_0^{2\pi} I_2(\theta,\theta') \cos(k\theta') \frac{ds}{d\theta}(\theta')\, d\theta' =: 
& \underbrace{\int_0^{2\pi} \left[I_2(\theta,\theta') - \frac{n_z(\theta)}{4\pi} \ln\sin^2\left(\frac{\theta-\theta'}{2}\right)\right]\cos(k\theta')\frac{ds}{d\theta}(\theta')\, d\theta'}_{\T_{2,k}(\theta)}  \\  
& + \frac{n_z(\theta)}{4\pi} \underbrace{ \int_0^{2\pi}\cos(k\theta')\ln\sin^2\left(\frac{\theta-\theta'}{2}\right) \frac{ds}{d\theta}(\theta')\, d\theta'}_{\Psi_k(\theta)},
\end{aligned}
\label{eq:intI2}
\end{equation}
where the regular part $\T_{2,k}(\theta)$ can be approximated with the
trapezoidal quadrature noting that
\begin{equation}
\lim_{\theta' \rightarrow \theta} \left[I_2(\theta,\theta') - \frac{n_z(\theta)}{4\pi} \ln\sin^2\left(\frac{\theta-\theta'}{2}\right)\right]
= - \Q(\theta),
\label{eq:intI20}
\end{equation}
where $\Q(\theta)$ was defined in \eqref{eq:intI10}.

The singular part $\Psi_k(\theta)$ can be evaluated using the
trigonometric series expansion \eqref{eq:dsdtN} for $ds/d\theta$ which
leads to
\begin{align}
\Psi_k(\theta) 
& = \int_0^{2\pi}\cos(k\theta')\ln\sin^2\left(\frac{\theta-\theta'}{2}\right) \frac{ds}{d\theta}(\theta')\, d\theta' \nonumber \\
& = \sum_{l=0}^{N} \eta_l \int_0^{2\pi}\cos(k\theta')\cos(l\theta')\ln\sin^2\left(\frac{\theta-\theta'}{2}\right) \, d\theta'   \nonumber \\
&\qquad + \zeta_l \int_0^{2\pi}\cos(k\theta')\sin(l\theta')\ln\sin^2\left(\frac{\theta-\theta'}{2}\right) \, d\theta'  \nonumber \\
& = \frac{1}{2} \sum_{l=0}^{N} \eta_l \left[C_{l+k}(\theta)+C_{l-k}(\theta)\right]  + \zeta_l \left[S_{l+k}(\theta)+S_{l-k}(\theta)\right],
\label{eq:psik}
\end{align}
where the last equality follows from the trigonometric identities
\begin{align*}
\cos(k\theta)\cos(l\theta) = \frac{1}{2}\left[ \cos\left[(l+k)\theta\right] + \cos\left[(l-k)\theta\right] \right], \\
\cos(k\theta)\sin(l\theta) = \frac{1}{2}\left[ \sin\left[(l+k)\theta\right] + \sin\left[(l-k)\theta\right] \right].
\end{align*}
The integrals in \eqref{eq:Djk} involving the sine functions are
approximated analogously.

Finally, the entries of matrix $\bE$ corresponding to the last term in
operator $\G$, are defined as follows
\begin{equation}
E_{jk} := \C\, 
\begin{cases}
-(k-1) \int_0^{2\pi} I_3(\theta_j,\theta')\sin\left[(k-1)\theta'\right] \, d\theta', & 1 \le k \le N+1 \\
(k-N-1) \int_0^{2\pi} I_3(\theta_j,\theta')\cos\left[(k-N-1)\theta'\right] \, d\theta', & N+1 < k \le M 
\end{cases},
\quad j=1,\dots,M,
\label{eq:Ejk}
\end{equation}
where as before the integrals can be evaluated using property
\eqref{eq:Ksing3} to split them into regular and singular parts. For
the integrals involving the sine functions in \eqref{eq:Ejk}, we thus
obtain when $k \ge 1$
\begin{equation}
\begin{aligned}
\int_0^{2\pi} I_3(\theta,\theta') \sin(k\theta') \, d\theta' =: 
& \underbrace{\int_0^{2\pi} \left[I_3(\theta,\theta') + \frac{r(\theta)}{4\pi} \ln\sin^2\left(\frac{\theta-\theta'}{2}\right)\right]\sin(k\theta')\, d\theta'}_{\T_{3,k}(\theta)}  \\  
& - \frac{r(\theta)}{4\pi} \underbrace{ \int_0^{2\pi}\sin(\theta')\ln\sin^2\left(\frac{\theta-\theta'}{2}\right) \, d\theta'}_{S_k(\theta)},
\end{aligned}
\label{eq:intI3}
\end{equation}
where the regular part $\T_{3,k}(\theta)$ can be approximated using
the trapezoidal quadrature noting that for $\theta \in [0,2\pi]$
\begin{equation}
\lim_{\theta' \rightarrow \theta} \left[I_3(\theta,\theta') + \frac{r(\theta)}{4\pi} \ln\sin^2\left(\frac{\theta-\theta'}{2}\right)\right]
= - \frac{r(\theta)}{4\pi} \left[ \ln\left(\frac{\frac{ds}{d\theta}}{4 r}\right)^2 + 4 \right],
\label{eq:intI30}
\end{equation}
whereas the singular part is given in \eqref{eq:Sk}. The integrals in
\eqref{eq:Ejk} involving the cosine functions are evaluated
analogously.

We now offer some comments about the validation of the approach
described above. The technique in which an improper integral is
approximated by splitting it into the regular and singular part,
cf.~\eqref{eq:intI0}, \eqref{eq:intI1}, \eqref{eq:intI2} and
\eqref{eq:intI3}, is standard \citep{h95} and was thoroughly tested on
similar problems by \citet{ep13,ProtasElcrat2016}. Here the
consistency of this approach was additionally validated by comparing
the results with the results obtained by approximating the integrals
on the LHS in \eqref{eq:intI0}, \eqref{eq:intI1}, \eqref{eq:intI2} and
\eqref{eq:intI3} with the trapezoidal quadrature based on a staggered
grid, such that the kernels were not evaluated at their singular
points, i.e., for $\theta' = \theta$. Finally, the ultimate validation
of the described numerical approach is provided by the rapid
convergence of Newton's method used to compute the equilibrium vortex
configurations, cf.~\S \ref{sec:rings}, where the Jacobian
\eqref{eq:DF} is constructed with the help of the matrices $\bB$,
$\bC$, $\bD$ and $\bE$, cf.~\eqref{eq:M}, used here to set up the
eigenvalue problem \eqref{eq:evalda} for the stability analysis. With
the high precision of the numerical quadratures thus established, the
shape differentiation results in \eqref{eq:vne_rhs} were validated by
comparing them against simple forward finite-difference approximations
of the shape derivatives.  For example, the consistency of the first
term on the RHS in \eqref{eq:vne_rhs} was checked by comparing it (as
a function of $\theta$) to
\begin{equation*}
\epsilon^{-1} \, \left( \n_{\x^{\epsilon}} - \n_{\x} \right) \cdot 
\left[\C \, \int_{\partial\A} \bK(\x(t),\x')\, ds_{\x'} - W\, \e_z\right]
\end{equation*}
in the limit of vanishing $\epsilon$. In the same spirit, the
consistency of the second and third term on the RHS of
\eqref{eq:vne_rhs} was verified by perturbing the evaluation point
$\x$ and the contour $\partial\A$, respectively.

Finally, we discuss solution of the constrained eigenvalue problem
\eqref{eq:evald}. We note that the problem needs to be restricted to
eigenvectors {satisfying condition} \eqref{eq:evaldb} and this is
done with a projection approach \citep{Golub1973} which will also
resolve the issue with system \eqref{eq:evald} being overdetermined.
Defining
\begin{equation}
\bM := \bB + \bC + \bD +  \bE,
\label{eq:M}
\end{equation}
system \eqref{eq:evalda} can be written as $\lambda \, \w = (-i
\bA^{-1}\, \bM) \, \w$. Condition \eqref{eq:evaldb} restricts
eigenvectors $\w$ to a subspace given by the kernel space $\N(\c)
{:= \{ \w \in \RR^M, \ \c\w = 0\}}$ of the constraint operator
$\c$ with the corresponding projection operator given by $\P_{\N(\c)}
:= \mathbf{I} - \c^{\dagger} \c$, where $\mathbf{I}$ is the $M\times
M$ identity matrix and $\c^{\dagger} := \c^T (\c\c^T)^{-1} = {\c^T /
  ||\c||_2}$ is the Moore-Penrose pseudo-inverse of the operator $\c$
\citep{l05}. {Applying this projection operator to the eigenvalue
  problem and noting that for vectors satisfying constraint
  \eqref{eq:evaldb} we have $\w = \P_{\N(\c)} \w$, we obtain $\lambda
  \P_{\N(\c)} \, \w = \lambda \,\w = \P_{\N(\c)} (-i \bA^{-1} \bM) \,
  \w$.  Next, employing} the identities {$\Sp\left(\P_{\N(\c)}
    \,(-i \bA^{-1} \bM)\right) = \Sp\left(\P_{\N(\c)}^2 \, (-i
    \bA^{-1} \bM)\right) = \Sp\left(\P_{\N(\c)}\, (-i \bA^{-1} \bM)
    \,\P_{\N(\c)}\right)$}, where $\Sp(\cdot)$ denotes the spectrum of
a matrix {and} which are a consequence of the properties of the
projection operator, we finally obtain an $M \times M$ algebraic
problem equivalent to \eqref{eq:evald} \citep{Golub1973}
\begin{equation}
\lambda \, \w = \P_{\N(\c)} (-i \bA^{-1} \bM) \P_{\N(\c)} \, \w 
\label{eq:Peval}
\end{equation}
which can be efficiently solved in MATLAB using the function {\tt
  eig}. Our numerical tests confirm the convergence of {$K$
  leading} eigenvalues obtained by solving problem \eqref{eq:Peval},
{where $1 \ll K < 2N$,} to well-defined limiting values as the
resolution $M$ is refined, with analogous behavior observed for the
corresponding eigenvectors $\w$. {For each resolution $M = 2N$, a
  certain number $(2N-K)$ of eigenvalues with the largest moduli are
  spurious, i.e., they are numerical artifacts related to
  discretization, which is a well-known effect \citep{b01}. In
  addition, since for $\alpha \rightarrow \sqrt{2}$ Norbury's vortices
  have to be determiend with a relaxed accuracy (cf.~\S
  \ref{sec:rings}), spurious eigenvalues with smaller magnitudes also
  appear for $\alpha \ge 1.35$ (such spurious eigenvalues are
  distinguished from the true ones by the fact that they do not
  exhibit the expected trends as the resolution $M$ is refined and/or
  when the parameter $\alpha$ is varied \citep{b01}). All of those
  spurious eigenvalues are excluded from the analysis in \S
  \ref{sec:comput}.}  {Based on such tests we can conservatively
  estimate the accuracy of the resolved eigenvalues as 8 significant
  digits for $\alpha \rightarrow 0$ and 3 significant digits when
  $\alpha \rightarrow \sqrt{2}$.}

\section{Computational Results}
\label{sec:comput}

In this section we present and analyze the results obtained by
numerically solving eigenvalue problem \eqref{eq:Peval}. Our objective
is to understand how the stability properties of the family of
Norbury's vortices, cf.~figure \ref{fig:vortices}(a), depend on the
parameter $\alpha$. In general, two distinct regimes can be observed:
when $\alpha \in (0,\alpha_0]$, where $\alpha_0 \approx 0.925$ is
determined approximately with the accuracy of $\Delta \alpha =
0.0125$, all eigenvalues are purely real, i.e., $\lambda_k =
\Re(\lambda_k)$, $k \in {[-K/2,\dots,K/2]}$, indicating that the
vortices are neutrally stable; on the other hand, when $\alpha \in
(\alpha_0, \sqrt{2})$, the spectra also include conjugate pairs of
eigenvalues with nonvanishing imaginary parts which correspond to
linearly stable and unstable eigenmodes.

\begin{figure}
\centering
\mbox{\hspace*{-1.2cm}
\subfigure[]{\includegraphics[width=0.575\textwidth]{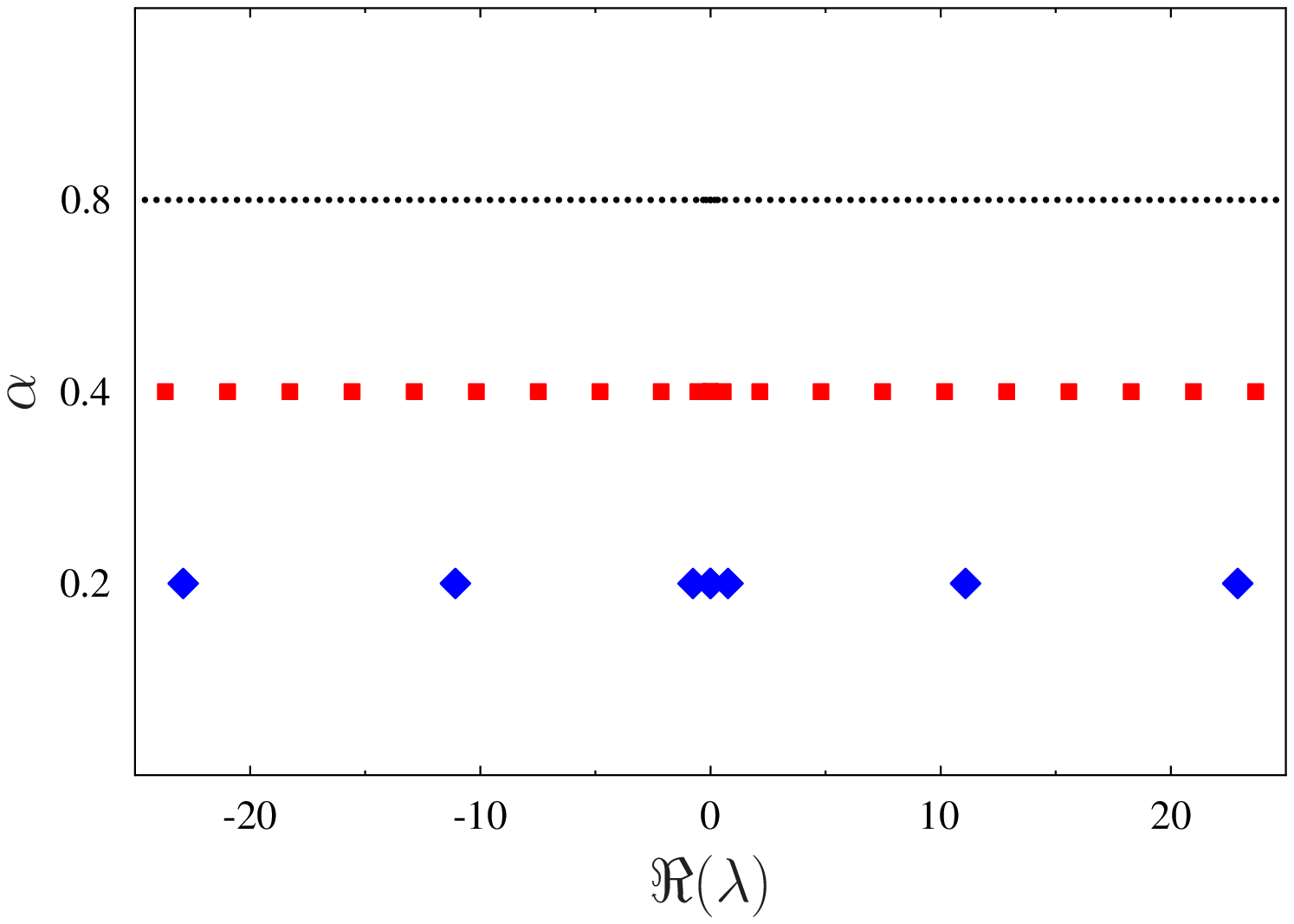}}
\subfigure[]{\includegraphics[width=0.575\textwidth]{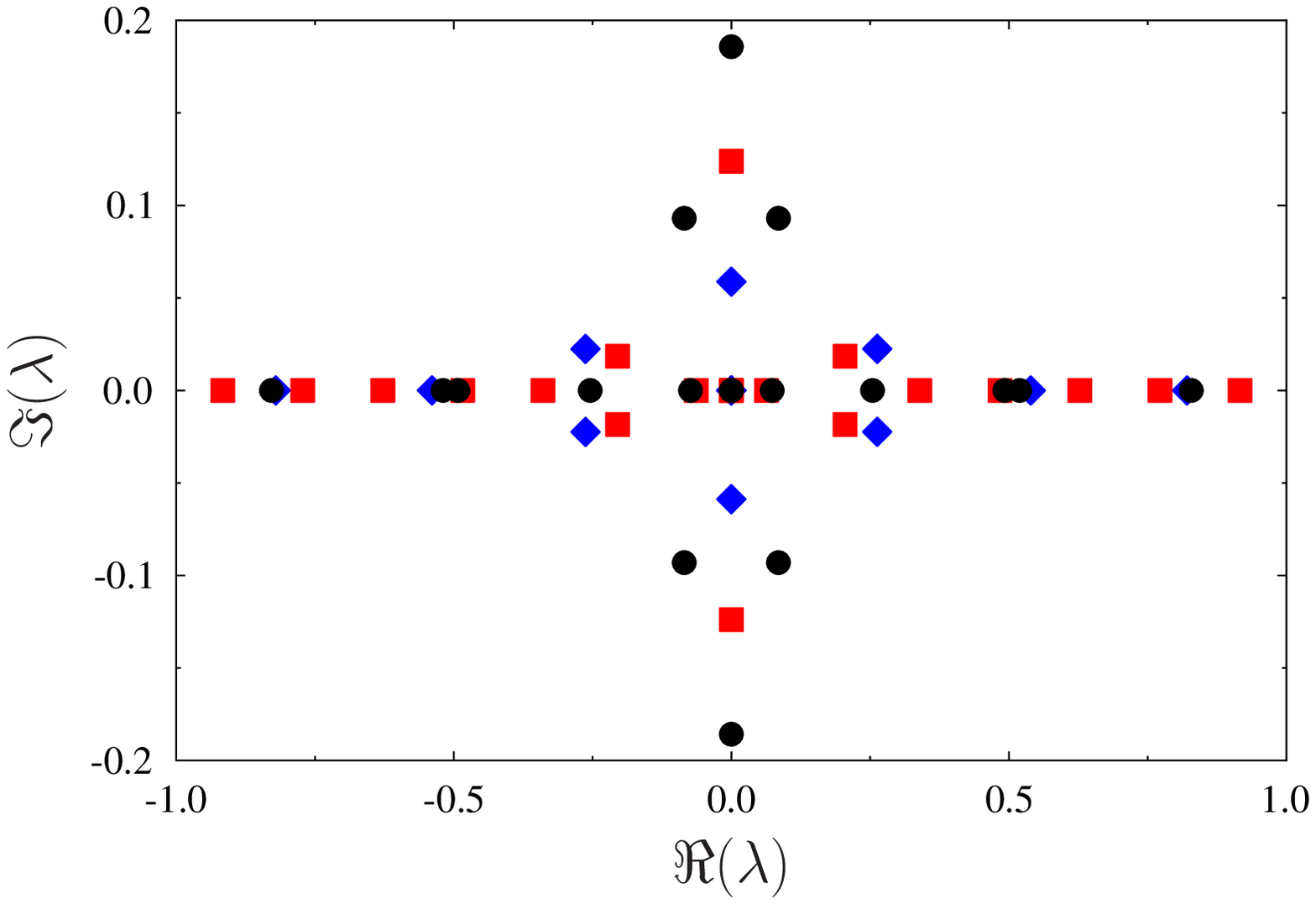}}}
\caption{Spectra of eigenvalue problem \eqref{eq:Peval} obtained for
  Norbury's vortices with (a) $\alpha = 0.2$ (blue diamonds),
  $\alpha=0.4$ (red squares), $\alpha=0.8$ (black circles) and (b)
  $\alpha = 0.975$ (blue diamonds), $\alpha=1.2$ (red squares),
  $\alpha=1.39$ (black circles). {In panel (a) all eigenvalues
    are purely real and, for clarity, the vertical axis is not scaled
    uniformly.}}
\label{fig:spectra}
\end{figure}

\begin{figure}
\centering
\mbox{\hspace*{-1.2cm}
\subfigure[]{\includegraphics[width=0.575\textwidth]{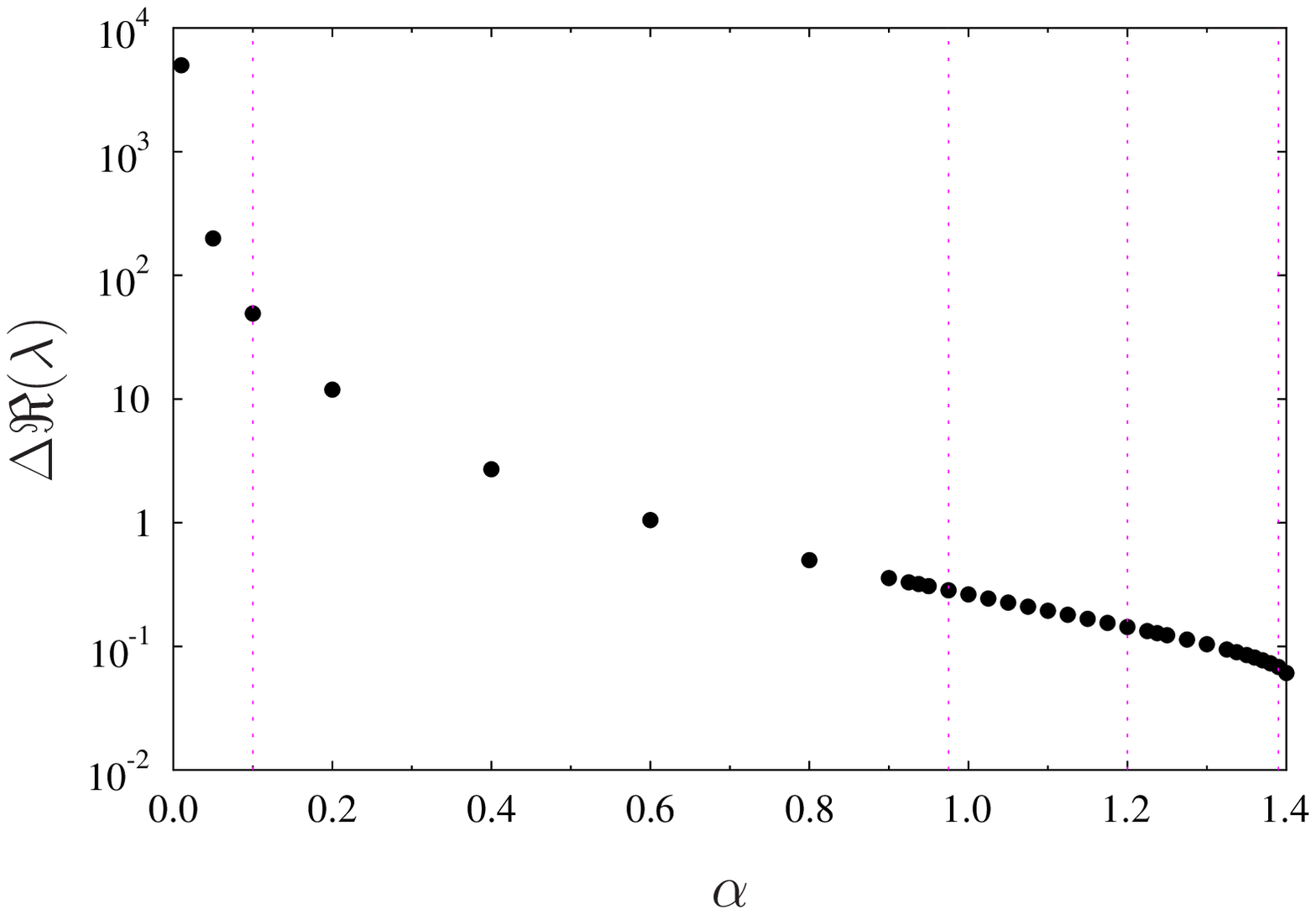}}
\subfigure[]{\includegraphics[width=0.575\textwidth]{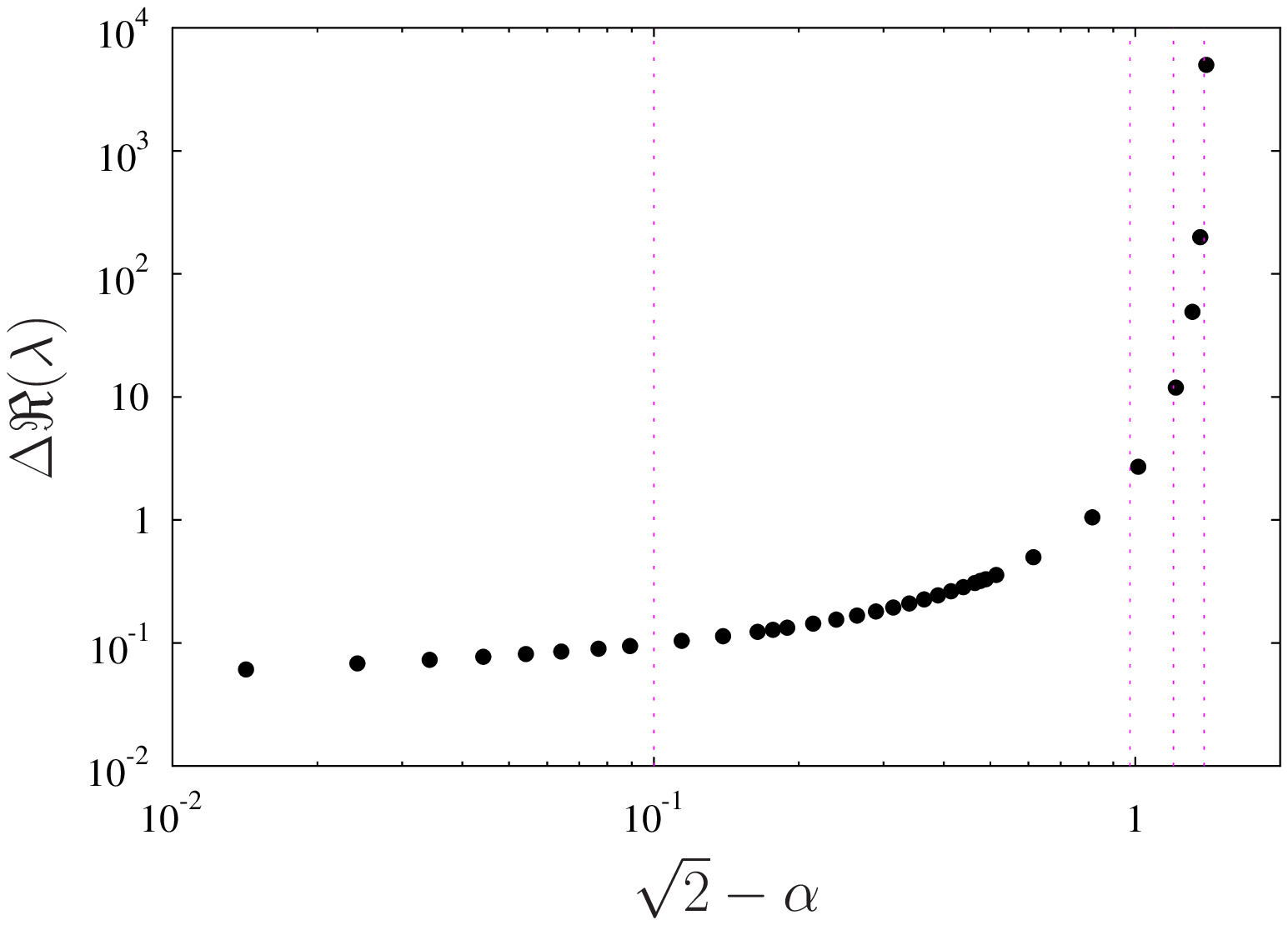}}}
\caption{Dependence of the average difference $\Delta\Re(\lambda)$
  between two consecutive purely real eigenvalues on (a) the parameter
  $\alpha$ with the linear scaling and (b) the quantity
  $(\sqrt{2}-\alpha)$ with the logarithmic scaling. The dotted vertical
  lines represent the values of $\alpha$ for which the unstable and
  certain neutrally stable eigenvectors are analyzed in figures
  \ref{fig:uevec1}, \ref{fig:uevec2} and \ref{fig:nevec}.}
\label{fig:evals_real}
\end{figure}

\begin{figure}

\centering
\includegraphics[width=0.575\textwidth]{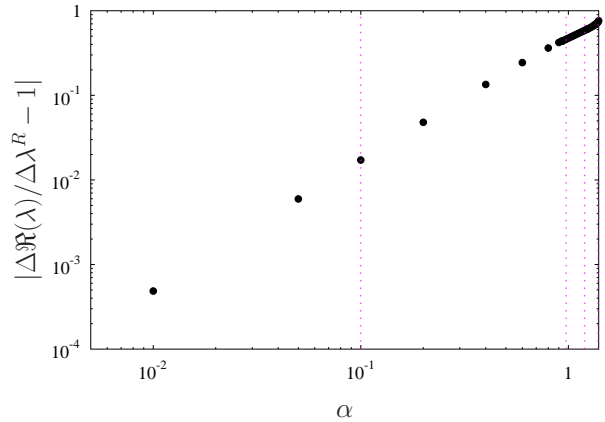}
\caption{{Dependence of the relative difference between the
    quantity $\Delta\Re(\lambda)$, cf.~figure \ref{fig:evals_real},
    and the quantity $\Delta\lambda^R$ characterizing the stability
    spectrum of Rankine's vortex on the parameter $\alpha$. The dotted
    vertical lines represent the values of $\alpha$ for which the
    unstable and certain neutrally stable eigenvectors are analyzed in
    figures \ref{fig:uevec1}, \ref{fig:uevec2} and \ref{fig:nevec}.}}
\label{fig:evals_kelvin}
\end{figure}

As regards the first regime with purely real eigenvalues, the spectra
obtained for $\alpha = 0.2, 0.4,0.8$ are presented in figure
\ref{fig:spectra}(a). As can be observed in this figure, the
difference $\Delta\Re(\lambda) = \lambda_{k+1} - \lambda_k$ between
two consecutive eigenvalues $\lambda_{k+1}$ and $\lambda_k$ (when they
are ordered such that $\Re(\lambda_{k+1}) > \Re(\lambda_k)$) is
approximately constant with respect to the index $k$, except for the
eigenvalues closest to the zero eigenvalue $\lambda_0 = 0$. It is also
evident from this figure that $\Delta\Re(\lambda)$ increases as
$\alpha$ is reduced. This effect is quantified in figure
\ref{fig:evals_real}(a), where we can see that $\Delta\Re(\lambda)
\rightarrow \infty$ in the thin-ring limit as $\alpha \rightarrow 0$.
{We now analyze the behavior of the eigenvalues obtained in the
  thin-ring limit in relation to} the stability properties of
Rankine's vortex, for which {the eigenvalues are given by
  $\lambda_k^R = \frac{\omega}{2}(k-1)$,} $k \in \ZZ$, where $\omega$
is the vorticity within the vortical region \citep{lamb-1932}.
{For Rankine's vortex the difference between two nearby
  eigenvalues is $\Delta\lambda^R := |\lambda_k^R - \lambda_{k-1}^R| =
  \frac{\omega}{2}$, where the vorticity corresponding to the
  thin-vortex limit of the vortex ring can be expressed as $\omega =
  \C \, r = 1 / \alpha^2$ (because as $\alpha \rightarrow 0$, $r
  \rightarrow L = 1$). The relative difference between the quantities
  $\Delta\Re(\lambda)$, cf.~figure \ref{fig:evals_real}, and
  $\Delta\lambda^R$ is shown as a function of $\alpha$ in figure
  \ref{fig:evals_kelvin} where we see that $\Delta\Re(\lambda)$
  approaches $\Delta\lambda^R$ faster than linearly in $\alpha$ as
  $\alpha$ vanishes (since the spectra of Rankine's vortex and of the
  thin vortex rings consist of uncountably many purely real
  eigenvalues, the present argument is more conveniently formulated in
  terms of average differences between nearby eigenvalues as done in
  figure \ref{fig:evals_kelvin}, rather than by analyzing the
  dependence of individual eigenvalues on $\alpha$). This demonstrates
  that the behavior of the eigenvalues in the thin-ring limit is in
  fact consistent with the stability properties of Rankine's vortex.
  In the opposite limit,} when $\alpha \rightarrow \sqrt{2}$, the
difference $\Delta\Re(\lambda)$ vanishes, albeit rather slowly,
cf.~figure \ref{fig:evals_real}(b), demonstrating that the purely real
eigenvalues become more densely packed in this limit and thus approach
a continuous spectrum, which is consistent with the stability
properties of Hill's vortex \citep{ProtasElcrat2016}.

\begin{figure}
\centering
\mbox{\hspace*{-1.2cm}
\subfigure[]{\includegraphics[width=0.575\textwidth]{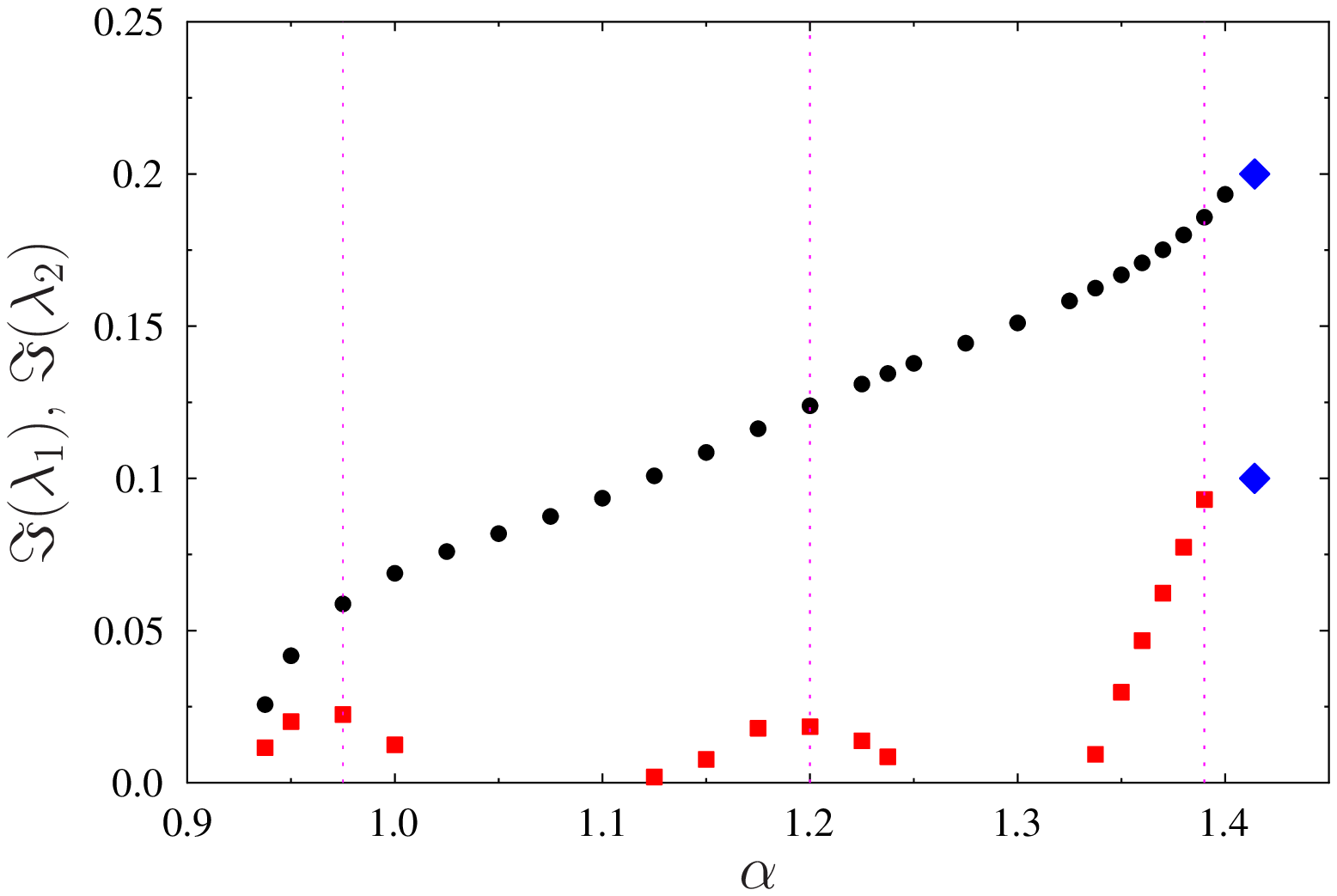}}
\subfigure[]{\includegraphics[width=0.575\textwidth]{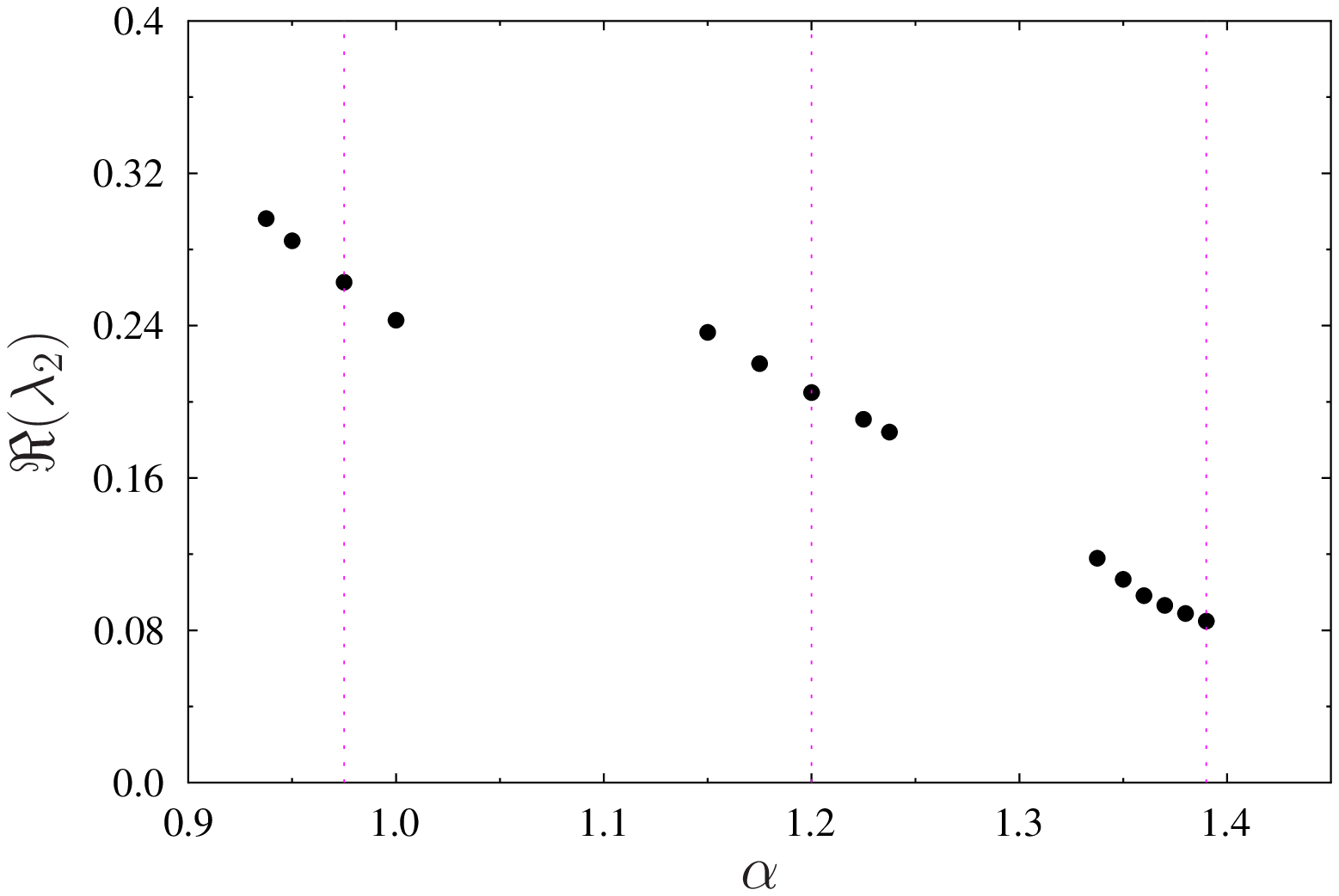}}}
\caption{Dependence of (a) the imaginary parts of the unstable
  eigenvalues $\lambda_1$ (black circles) and $\lambda_2$ (red
  squares) and (b) the real part of the unstable eigenvalue
  $\lambda_2$ on the parameter $\alpha$. The blue diamonds in panel
  (a) represent the eigenvalues characterizing the stability of Hill's
  vortex {\citep{mm78,ProtasElcrat2016}}. Since due to the
  reasons discussed {at the end of \S \ref{sec:numer}} the
  eigenvalue $\lambda_2$ could not be reliably determined when $\alpha
  = 1.4$, it is not shown in these figures. The dotted vertical lines
  represent the values of $\alpha$ for which the unstable eigenvectors
  are analyzed in figures \ref{fig:uevec1} and \ref{fig:uevec2}.}
\label{fig:evals_cplx}
\end{figure}

Concerning the second regime in which linearly stable and unstable
eigenmodes appear, representative spectra are shown in figure
\ref{fig:spectra}(b). In all cases we note the presence of a conjugate
pair of purely imaginary eigenvalues, which we will denote $\lambda_1$
and $\overline{\lambda}_1$, and two conjugate pairs of eigenvalues
with nonzero real parts of equal magnitude, which we will denote
$\lambda_2$, $\overline{\lambda}_2$, $-\lambda_2$,
$-\overline{\lambda}_2$, in addition to a large number of uniformly
spaced purely real eigenvalues.  As regards the dependence of the
imaginary parts of the eigenvalues $\lambda_1$ and $\lambda_2$ on
$\alpha$, in figure \ref{fig:evals_cplx}(a) we observe that for all
$\alpha \in (\alpha_0, \sqrt{2})$ we have $|\Im(\lambda_1)| >
|\Im(\lambda_2)|$, indicating that the purely imaginary eigenvalue
$\lambda_1$ is associated with the most unstable eigenmode. In figure
\ref{fig:evals_cplx}(a) we also note that $\Im(\lambda_1)$ grows
{monotonically} as $\alpha$ increases and $\lim_{\alpha
  \rightarrow \sqrt{2}} \lambda_1 = i0.2$, which is equal to the
eigenvalue associated with the most unstable eigenmode of Hill's
vortex \citep{mm78,ProtasElcrat2016}. Concerning the dependence of
$|\Im(\lambda_2)|$ on $\alpha$, we note that, interestingly, it is not
{monotonic} and in fact the unstable eigenvalues $\lambda_2$ are
present only when $\alpha \in [\alpha_0,\alpha_1] \cup
[\alpha_2,\alpha_3] \cup [\alpha_4,\sqrt{2})$, where $\alpha_1 \approx
1.0$, $\alpha_2 \approx 1.125$, $\alpha_3 \approx 1.2375$ and
$\alpha_4 \approx 1.3375$ are determined approximately with the
precision of $\Delta \alpha = 0.0125$. This demonstrates that the
associated eigenmodes are unstable only for certain values of
$\alpha$. In figure \ref{fig:evals_cplx}(b) we see that
$\Re(\lambda_2)$ decreases with $\alpha$ whenever $\Im(\lambda_2) \neq
0$. In particular, we note that $\lim_{\alpha \rightarrow \sqrt{2}}
\lambda_2 = i0.1$, which is the eigenvalue associated with the second
unstable eigenmode of Hill's vortex \citep{ProtasElcrat2016}. We thus
conclude that the eigenvalue spectra of Norbury's vortex rings
approach the spectra of Rankine's vortex and Hill's spherical vortex,
respectively, in the limits $\alpha \rightarrow 0$ and $\alpha
\rightarrow \sqrt{2}$ corresponding to the thin and fat vortex rings.


\begin{figure}
\centering
\mbox{\hspace*{-1.2cm}
\subfigure[]{\includegraphics[width=0.575\textwidth]{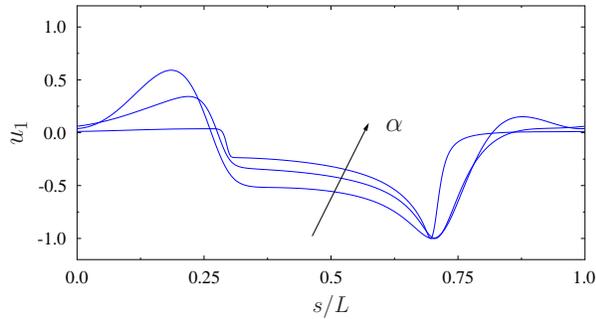}}
\subfigure[$\alpha = 0.975$]{\includegraphics[width=0.575\textwidth]{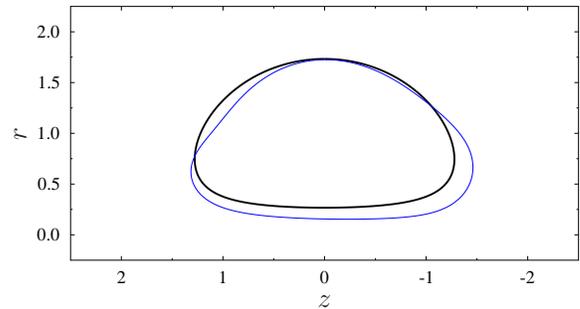}}}
\mbox{\hspace*{-1.2cm}
\subfigure[$\alpha = 1.2$]{\includegraphics[width=0.575\textwidth]{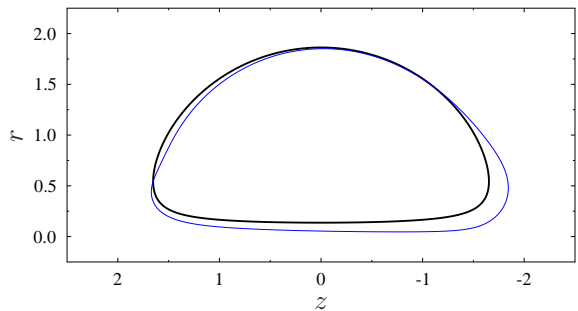}}
\subfigure[$\alpha = 1.39$]{\includegraphics[width=0.575\textwidth]{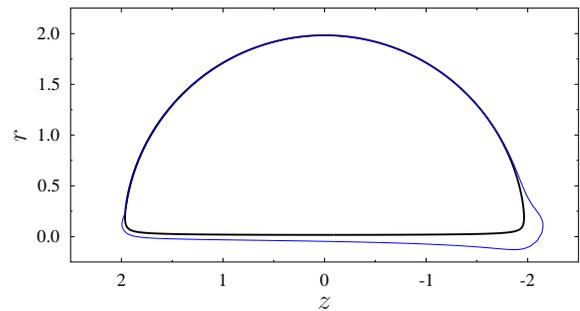}}}
\caption{(a) Unstable {purely real} eigenvectors $u_1$ obtained
  for $\alpha = 0.975, 1.2, 1.39$ presented as functions of the
  normalized arclength coordinate $s / L$ along the vortex boundary
  $\partial A$ (the arrow indicates the trend when $\alpha$
  increases). (b--d) Deformations of the vortex boundary induced by
  the eigenvectors $u_1$ for the indicated values of $\alpha$ (the
  magnitude of the perturbation, given by $\epsilon$ in ansatz
  \eqref{eq:xeps2}, is arbitrary). {The thick lines represent the
    vortex boundaries $\partial\A$ in their equilibrium
    configurations.}  In this and in subsequent figures the
  eigenvectors are scaled such that {$\max_{s\in[0,L]} |u(s)| =
    1$.} }
\label{fig:uevec1}
\end{figure}
\begin{figure}

\centering
\mbox{\hspace*{-1.2cm}
\subfigure[]{\includegraphics[width=0.575\textwidth]{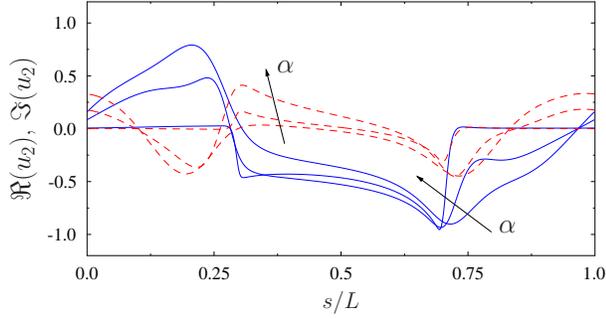}}
\subfigure[$\alpha = 0.975$]{\includegraphics[width=0.575\textwidth]{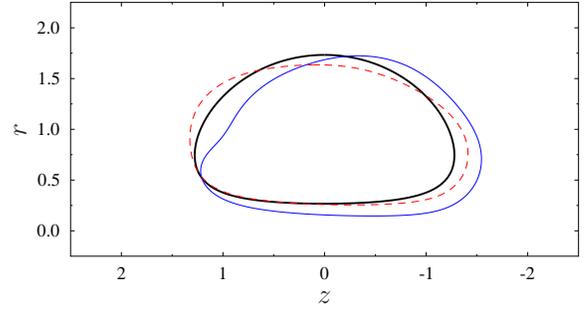}}}
\mbox{\hspace*{-1.2cm}
\subfigure[$\alpha = 1.2$]{\includegraphics[width=0.575\textwidth]{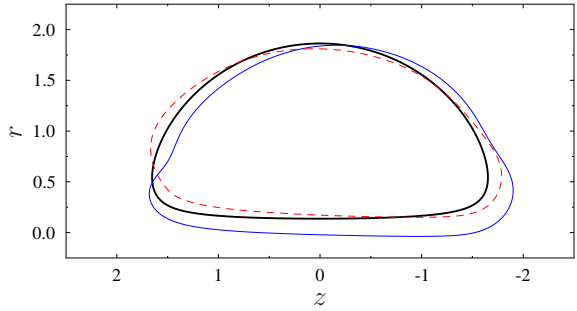}}
\subfigure[$\alpha = 1.39$]{\includegraphics[width=0.575\textwidth]{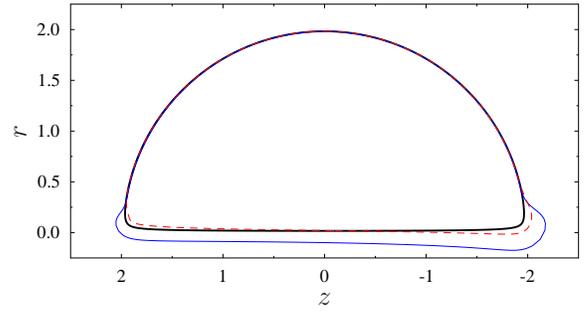}}}
\caption{(a) Real parts (blue solid lines) and imaginary parts (red
  dashed lines) of the unstable eigenvectors $u_2$ obtained for
  $\alpha = 0.975, 1.2, 1.39$ presented as functions of the normalized
  arclength coordinate $s / L$ along the vortex boundary $\partial A$
  (the arrow indicates the trend when $\alpha$ increases). (b--d)
  Deformations of the vortex boundary induced by the real and
  imaginary parts of the eigenvectors $u_2$ for the indicated values
  of $\alpha$ (the magnitude of the perturbation, given by $\epsilon$
  in ansatz \eqref{eq:xeps2}, is arbitrary). {The thick lines represent
  the vortex boundaries $\partial\A$ in their equilibrium
  configurations.}}
\label{fig:uevec2}
\end{figure}

\begin{figure}
\centering
\mbox{\hspace*{-1.2cm}
\subfigure[]{\includegraphics[width=0.575\textwidth]{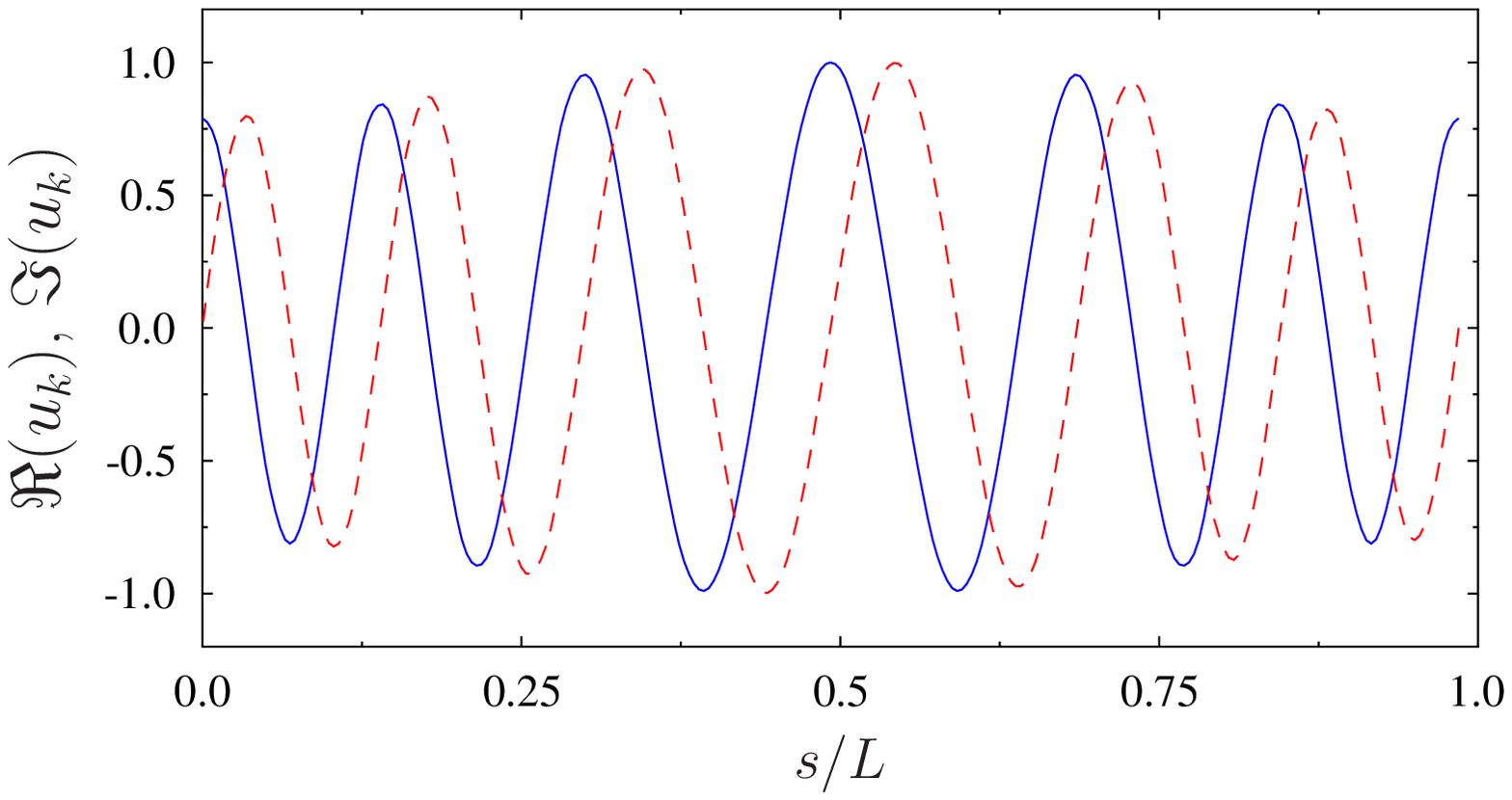}}
\subfigure[]{\includegraphics[width=0.575\textwidth]{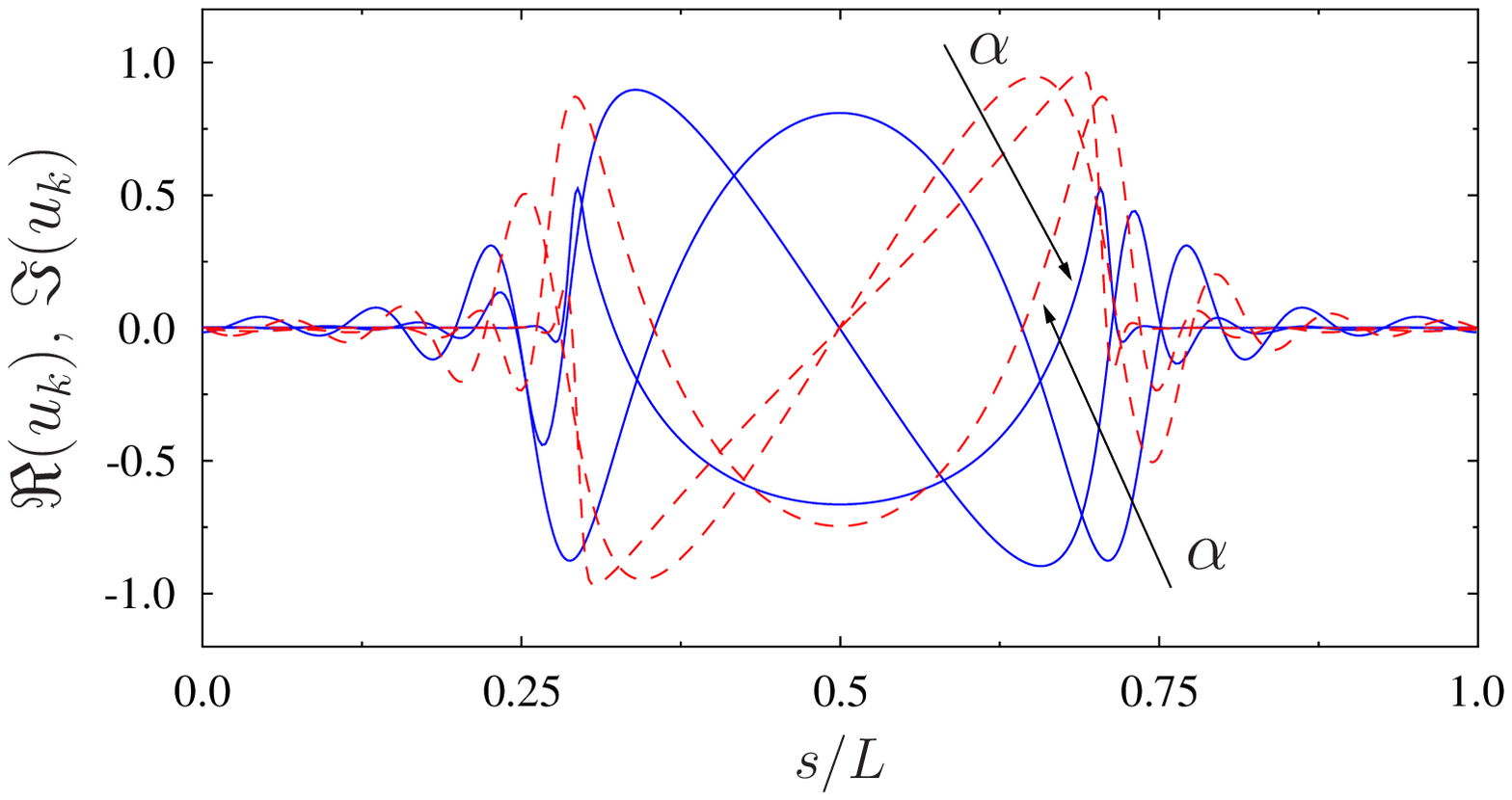}}}
\mbox{\hspace*{-1.2cm}
\subfigure[$\alpha = 0.1$]{\includegraphics[width=0.575\textwidth]{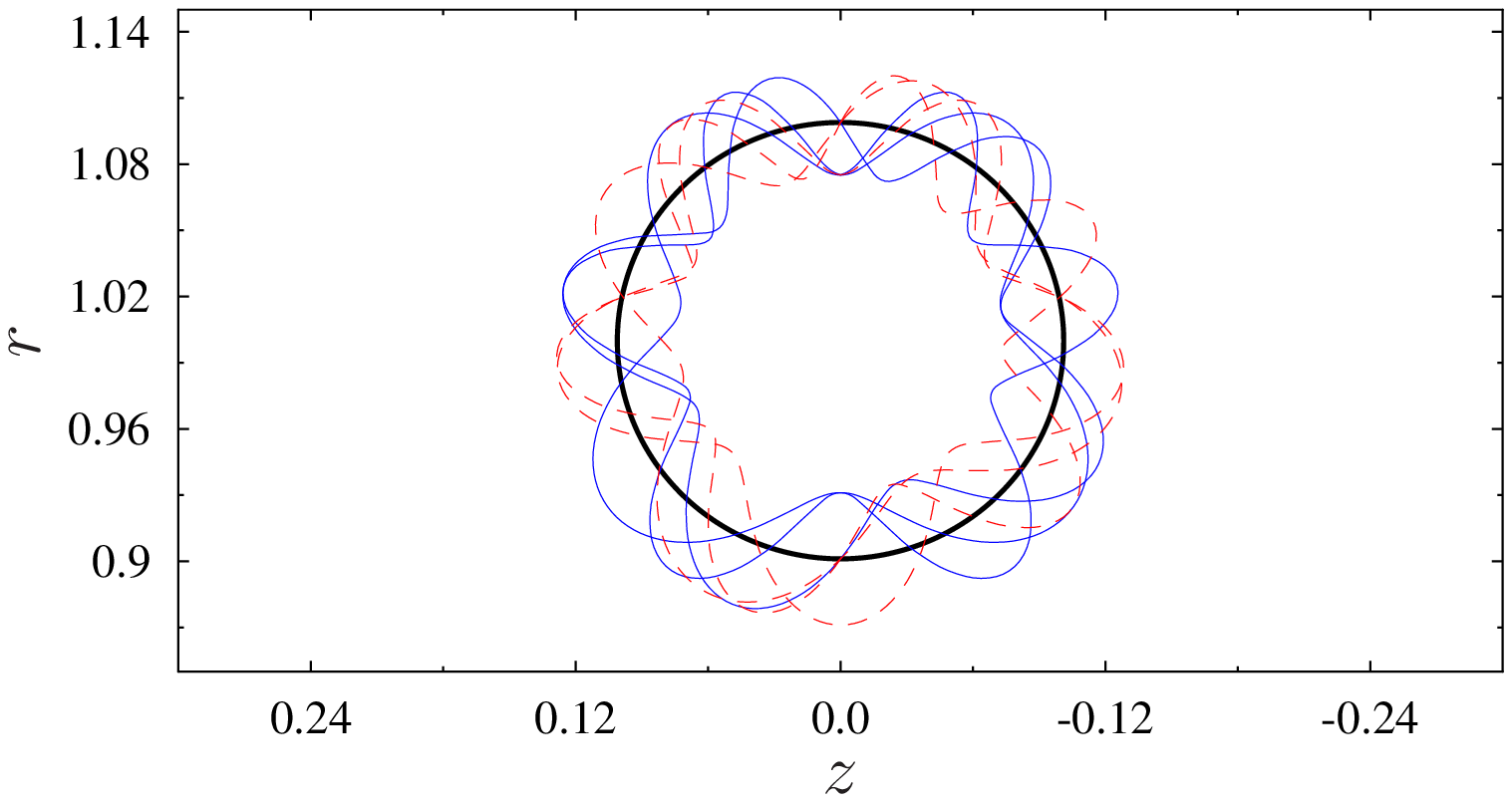}}
\subfigure[$\alpha = 0.975$]{\includegraphics[width=0.575\textwidth]{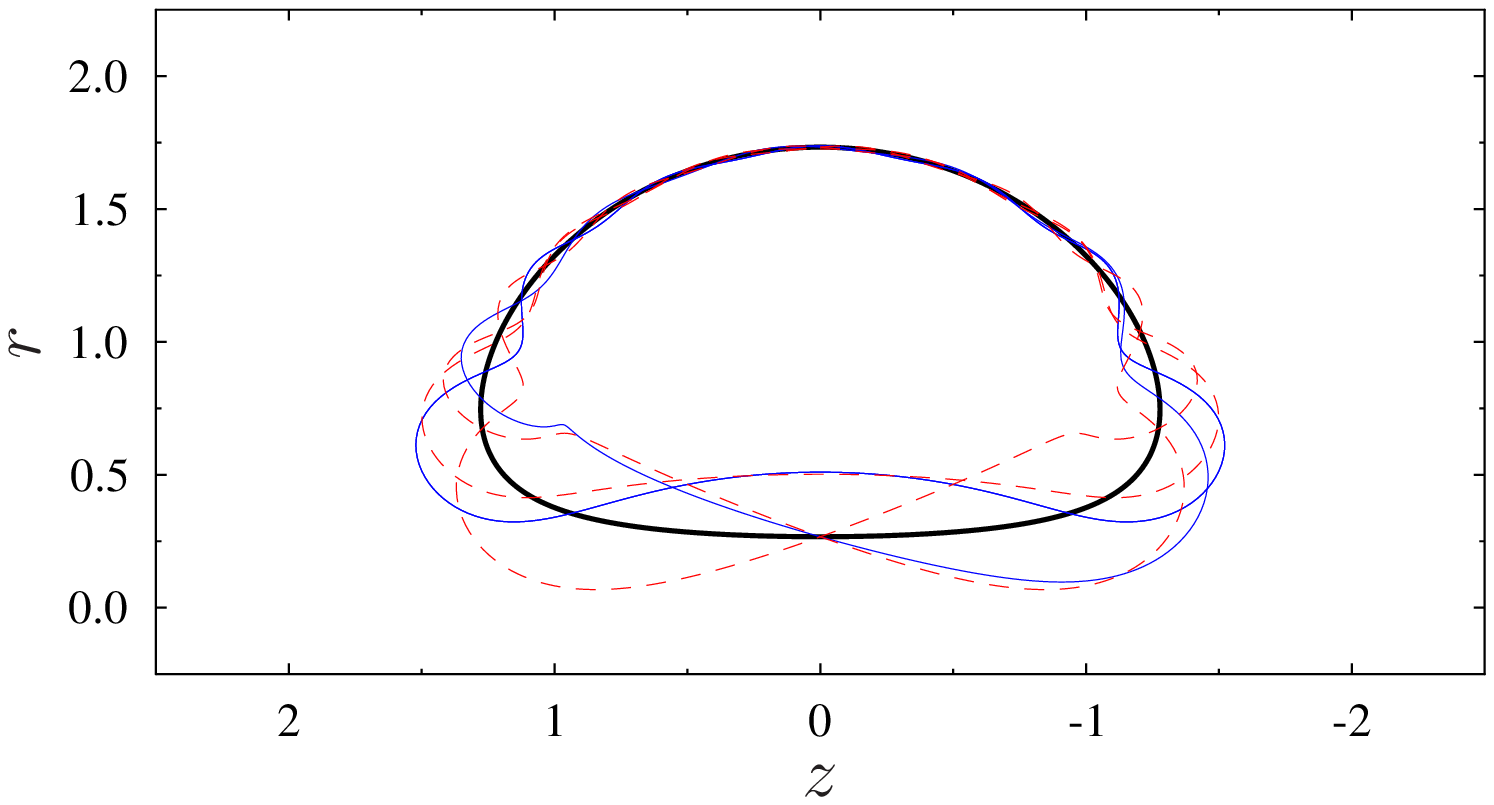}}}
\mbox{\hspace*{-1.2cm}
\subfigure[$\alpha = 1.2$]{\includegraphics[width=0.575\textwidth]{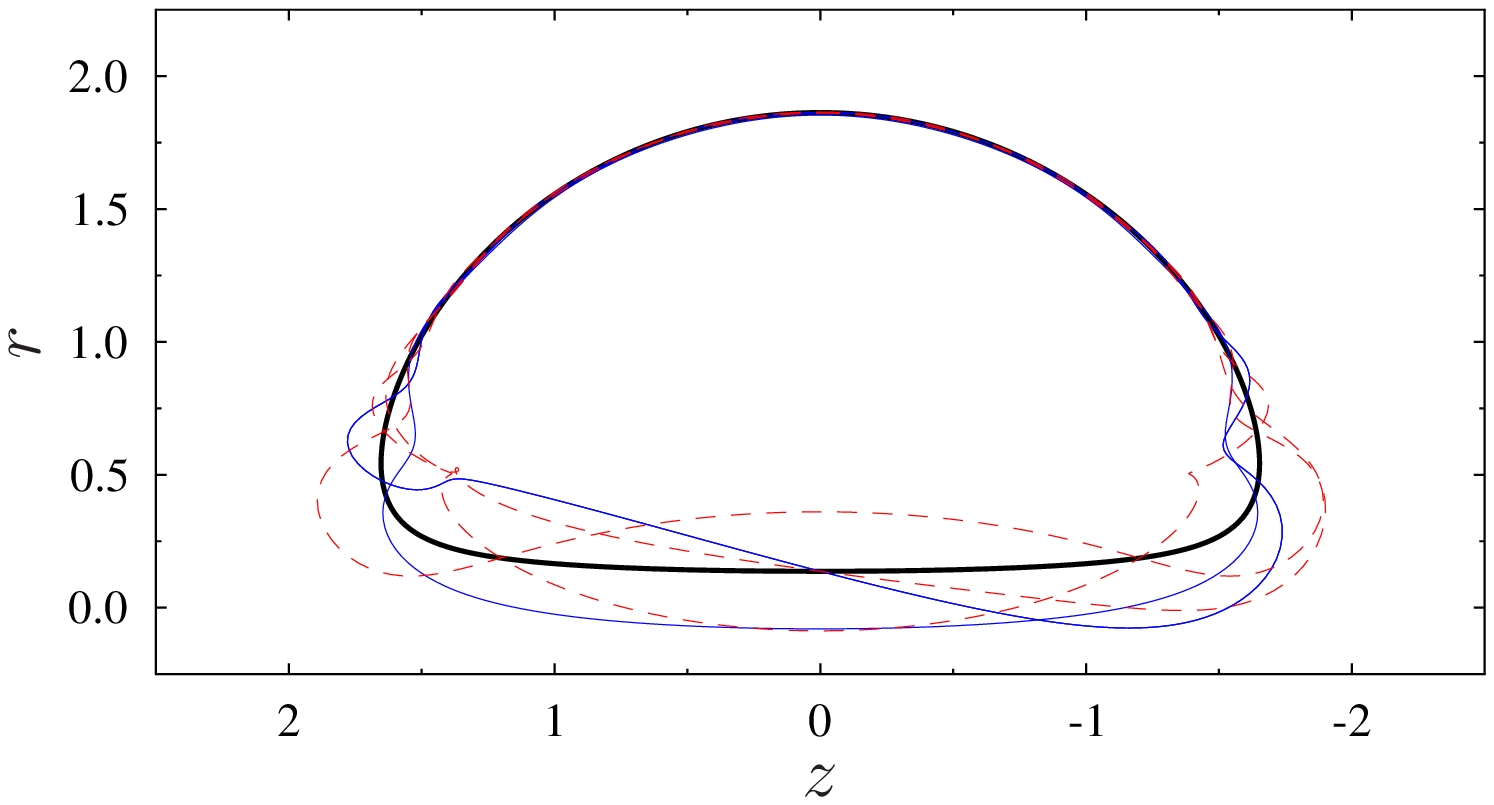}}
\subfigure[$\alpha = 1.39$]{\includegraphics[width=0.575\textwidth]{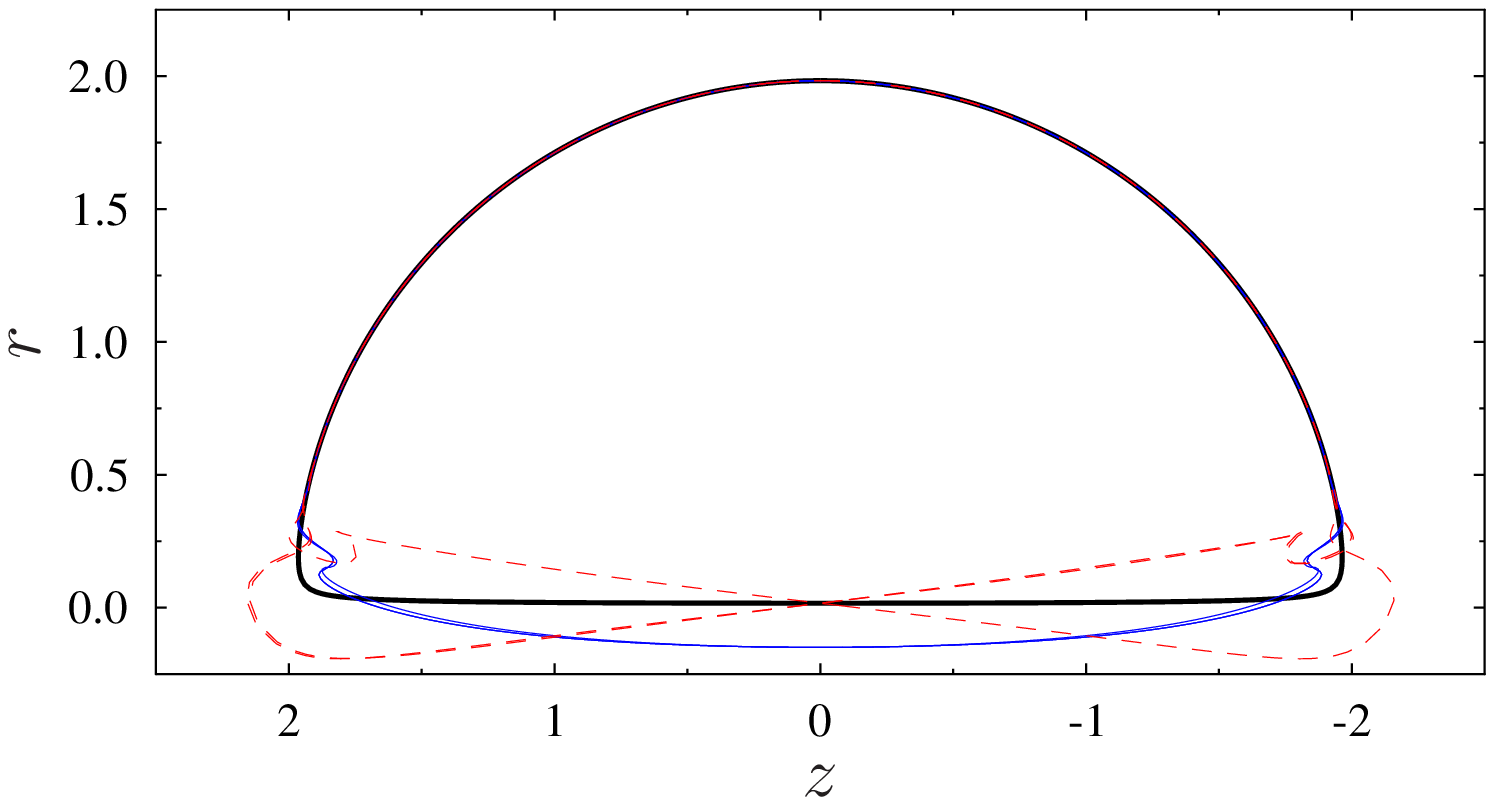}}}
\caption{Real parts (blue solid lines) and imaginary parts (red dashed
  lines) of representative neutrally stable eigenvectors $u_k$
  obtained for (a) $\alpha = 0.1$ and (b) $\alpha = 0.975, 1.2, 1.39$
  presented as functions of the normalized arclength coordinate $s /
  L$ along the vortex boundary $\partial A$ (the arrows in panel (b)
  indicate the trends when $\alpha$ increases). The eigenvectors are
  associated with purely real eigenvalues $\lambda_k$ close to $\pm
  0.05$.  (c--f) Deformations of the vortex boundary induced by the
  real and imaginary parts of these neutrally stable eigenvectors for
  the indicated values of $\alpha$ (the magnitude of the perturbation,
  given by $\epsilon$ in ansatz \eqref{eq:xeps2}, is arbitrary).
  {The thick lines represent the vortex boundaries $\partial\A$
    in their equilibrium configurations.}}
\label{fig:nevec}
\end{figure}

We now move on to discuss the eigenvectors of problem \eqref{eq:evalp}
focusing on how their structure depends on the parameter $\alpha$. We
begin with the unstable eigenvectors corresponding to the eigenvalues
$\lambda_1$ and $\lambda_2$, denoted $u_1$ and $u_2$, which we will
consider for representative values of the parameter $\alpha$ chosen
from each of the three subintervals of $\alpha$ where $\Im(\lambda_2)
\neq 0$, cf.~figure \ref{fig:evals_cplx}(a). We remark that when
ansatz \eqref{eq:r} is used and the eigenvalue $\lambda$ turns out to
be purely imaginary, then the imaginary part of the corresponding
eigenvector $u$ is irrelevant for the stability analysis (in fact, in
the present computations it vanishes identically in such cases). On
the other hand, if the eigenvalue $\lambda$ is complex, then both the
real and imaginary part of the eigenvector $u$ play a role in the
stability analysis. {We add that purely imaginary eigenvalues
  correspond to exponential in time growth of the normal perturbations
  $\rho(t,s)$ to the vortex boundary $\partial \A$, cf.~ans\"atze
  \eqref{eq:xeps2} and \eqref{eq:r}, whereas for complex eigenvalues
  this exponential growth is combined with harmonic oscillations.}

The unstable eigenvectors $u_1$ obtained for $\alpha = 0.975, 1.2,
1.39$ are compared as functions of the normalized arclength coordinate
$s / L$ measured counterclockwise along the vortex boundary $\partial
A$ in figure \ref{fig:uevec1}(a) (the arclength coordinate is measured
starting from the outer intersection of the vortex boundary $\partial
A$ with the $r$-axis, cf.~$\theta = 0$ in figure \ref{fig:A}, whereas
$L := \int_{\partial A}\, ds$ is the length of the vortex boundary).
Deformations of the vortex boundary $\partial A$ induced by these
eigenvectors are shown for the different values of $\alpha$ in figures
\ref{fig:uevec1}(b--d). We note that for all considered values of
$\alpha$ the deformation of the vortex boundary is concentrated near
its rear {(right)} extremity and becomes more localized as
$\alpha$ increases which is accompanied by the formation of a
high-curvature region in the vortex boundary. We also observe that
away from the rear extremity of the vortex the eigenvectors $u_1$
deform the boundary $\partial A$ in an opposite sense, which is a
consequence of the constraint requiring the perturbations to preserve
the circulation, cf.~\eqref{eq:r0}. The real and imaginary parts of
the eigenvectors $u_2$ associated with the second eigenvalue
$\lambda_2$ obtained for different values of $\alpha$ are shown in
figure \ref{fig:uevec2}(a) as functions of the normalized arclength
coordinate $s / L$, with the corresponding deformations of the vortex
boundary $\partial A$ illustrated in figures \ref{fig:uevec2}(b--d).
Here, for both the real and imaginary parts of the eigenvectors we
observe properties qualitatively similar to what was already noted for
the eigenvectors $u_1$ in figure \ref{fig:uevec1} with the
eigenvectors $u_2$ also becoming increasingly localized near the rear
extremity of the vortex as $\alpha$ increases. We emphasize that this
behavior is indeed consistent with the stability properties of Hill's
vortex, corresponding to $\alpha = \sqrt{2}$, where the unstable
eigenvectors $u_1$ and $u_2$ were found to have the form of singular
distributions localized at the rear stagnation point
\citep{ProtasElcrat2016}.  {The unstable eigenvectors $u_2$ also
  have the interesting property that while their real and imaginary
  parts deform the equilibrium contour in the same sense near the rear
  extremity of the vortex ring, they do so in an opposite sense near
  the front extremity.}  We add that since the eigenvalues with
nonzero imaginary parts come in conjugate pairs, cf.~figure
\ref{fig:spectra}(b), for each unstable mode there is a corresponding
stable mode. These stable eigenvectors have the form $u_1(L-s)$ and
$u_2(L-s)$ for $s \in [0,L]$, i.e., they are obtained from the
unstable modes by reflecting them with respect to the symmetry axis of
the vortex region $A$, such that they become localized near its front
{(left)} extremity as $\alpha \rightarrow \sqrt{2}$.

Finally, we move on to discuss the dependence of the neutrally stable
eigenvectors, denoted $u_k$, $k \neq 1,2$, associated with the purely
real eigenvalues on the parameter $\alpha$. {Extrapolating the
  data shown in figure \ref{fig:spectra}(a), we note that for all
  values of $\alpha \in (0,\sqrt{2})$ there is a countable infinity of
  neutrally-stable eigenvectors,} although only a finite subset of
these eigenvectors may be resolved with computations performed with a
finite resolution $M$. {Thus,} to fix attention, we will focus
here on a few representative eigenvectors associated with eigenvalues
$\lambda_k$ close to $\pm 0.05$. The real and imaginary parts of these
eigenvectors obtained for Norbury's vortices with $\alpha = 0.1$ and
$\alpha = 0.975, 1.2, 1.39$ are shown as functions of the normalized
arclength coordinate $s / L$ along the vortex boundary $\partial A$ in
figures \ref{fig:nevec}(a) and \ref{fig:nevec}(b), respectively,
whereas the deformations of the vortex boundaries induced by these
eigenvectors are illustrated in figures \ref{fig:nevec}(c--f). In
figures \ref{fig:nevec}(a) and \ref{fig:nevec}(c) we see that for
small values of $\alpha$, when the vortex region $A$ is nearly
circular, cf.~figure \ref{fig:vortices}(a), the eigenvectors have the
form of a nearly harmonic oscillation with a small modulation and with
the real and imaginary parts differing only by a phase shift. In fact,
in the thin-vortex limit $\alpha \rightarrow 0$ these eigenvectors
approach Kelvin waves (i.e., purely harmonic oscillations) known to
characterize the neutral stability of Rankine's vortex
\citep{lamb-1932}. On the other hand, for $\alpha \rightarrow
\sqrt{2}$ the neutrally stable eigenvectors have the form of
oscillations which are increasingly localized near the front
{(left)} and rear {(right)} extremity of the vortex region,
a behavior which in the limit is also consistent with the stability
properties of Hill's vortex \citep{ProtasElcrat2016}. In general, for
a given $\alpha$ the number of oscillations exhibited by the neutrally
stable eigenvectors increases with $|\Re(\lambda_k)|$.

\section{Summary and Conclusions}
\label{sec:final}

In this final section we briefly summarize and discuss our results
before pointing to some open questions. Recognizing that Norbury's
vortex rings are solutions to the {\em free-boundary} problem
\eqref{eq:Euler3D}--\eqref{eq:f}, their stability to axisymmetric
perturbations was studied here using an approach based on the
shape-differential calculus which allows one to rigorously account for
the effect of perturbations to the shape of the vortex boundary in the
linearization of the governing equations, cf.~\S \ref{sec:stab}.
{In addition to axisymmetry, the perturbations are assumed not
  the change the circulation $\Gamma$ of the vortex ring from its
  equilibrium configuration. Such perturbations may be therefore
  regarded as resulting from the application of conservative forces to
  the vortex ring.}  Information about the stability of the vortex
rings {to such axisymmetric and circulation-preserving
  perturbations} is then encoded in the spectrum of a singular
integro-differential operator \eqref{eq:La} defined on the vortex
boundary $\partial A$ in the meridional plane.  After discretizing it
with a spectrally-accurate numerical technique and imposing the
constraint ensuring that perturbations preserve the circulation of the
vortex ring, we obtain a constrained eigenvalue problem which is
solved numerically. Developed and validated by \citet{ep13}, this
approach was used earlier to solve an analogous stability problem for
Hill's vortex \citep{ProtasElcrat2016}. The base states (Norbury's
vortex rings) were recomputed using Newton's method based on
shape-differentiation to ensure that they are determined with a much
higher accuracy and for a broader range of the parameter $\alpha$ than
in Norbury's original work \citep{norbury-1973-JFM}.  The data for
these base states is provided as Supplementary Information
accompanying this paper.

The results presented in \S \ref{sec:comput} demonstrate that in terms
of stability properties {with respect to axisymmetric
  perturbations} the continuous family of Norbury's vortex rings
represents a ``bridge'' connecting Rankine's columnar vortex with
Hill's spherical vortex. Indeed, for small $\alpha \in (0, \alpha_0]$,
Norbury's vortex rings are neutrally stable with the eigenvectors
{and eigenvalues approaching Kelvin waves and their eigenvalues}
in the thin-vortex limit $\alpha \rightarrow 0$. The fact that the
stability properties of Rankine's vortex are recovered in this limit
can be explained by the observation that in this case a thin vortex
ring can be treated as an isolated columnar (Rankine) vortex and under
the assumption of axisymmetry perturbations are allowed in the
meridional plane only where they give rise to Kelvin waves. As
$\alpha$ increases and the vortex rings grow fatter, they become
unstable at $\alpha_0 \approx 0.925$ which is marked by the emergence
of two unstable eigenmodes. {There seems to be no obvious change
  to the form of the vortex region $\A$ or to the corresponding flow
  pattern to which this change of stability could be attributed.
  However, as observed by \citet{Ofarrell_dabiri_2012}, the change of
  stability can be explained by the fact that as $\alpha$ increases
  parts of the vortex boundary approach the rear stagnation point
  (which is on the flow axis) such that the resulting shear becomes
  strong enough to amplify the perturbation.  On the other hand, for
  thin vortex rings corresponding to small values of $\alpha$, the
  shear affecting the vortex boundaries is rather weak and
  perturbations are swept along the vortex perimeter without
  amplification.}  The most unstable mode $u_1$ is associated with a
purely imaginary eigenvalue $\lambda_1$ whose imaginary part increases
in magnitude {monotonically} with $\alpha$, cf.~figure
\ref{fig:evals_cplx}(a). Interestingly, the second unstable eigenmode
$u_2$ is present only for certain values of the parameter $\alpha \in
[\alpha_0, \sqrt{2})$. It is associated with the eigenvalue
$\lambda_2$ with a nonvanishing real part whose magnitude deceases
with $\alpha$, cf.~figure \ref{fig:evals_cplx}(b).  As a result, in
the fat-vortex limit $\alpha \rightarrow \sqrt{2}$ the two unstable
eigenvalues $\lambda_1$ and $\lambda_2$ converge to the eigenvalues
associated with the two unstable eigenvectors of Hill's vortex, while
the corresponding eigenvectors $u_1$ and $u_2$ become increasingly
localized near the rear extremity of the vortex. In the limit $\alpha
\rightarrow \sqrt{2}$ this high-curvature region becomes the rear
stagnation point of Hill's vortex. The two unstable eigenmodes of
Hill's vortex have the form of singular distributions localized at
that point \citep{ProtasElcrat2016} and can be therefore regarded as
the limits the unstable eigenvectors $u_1$ and $u_2$ of Norbury's
vortices converge to as $\alpha \rightarrow \sqrt{2}$, cf.~figures
\ref{fig:uevec1}(a) and \ref{fig:uevec2}(a). {We add that
  analogous results concerning the stability of Hill's vortex to
  axisymmetric circulation-preserving perturbations had been
  {earlier} obtained using different techniques by \citet{mm78}.}
At the same time, the set of purely real eigenvalues converges to the
continuous spectrum which was also reported for Hill's vortex,
although the convergence is rather slow, cf.~figure
\ref{fig:evals_real}(b).  Finally, we add that the properties of the
stability spectra obtained here for Norbury's vortex rings with
different values of $\alpha$, cf.~figures \ref{fig:spectra}(a) and
\ref{fig:spectra}(b), are consistent with the Hamiltonian structure of
the governing Euler equations, {in the sense that in all cases
  the sum of the eigenvalues is zero.}

{The stability results summarized above are consistent with the
  responses of Norbury's vortex rings to axisymmetric perturbations
  computed in the nonlinear regime by \citet{Ofarrell_dabiri_2012},
  see also \citet{YeChu1995}. More specifically, for thin vortex rings
  \citet{Ofarrell_dabiri_2012} found that small perturbations to the
  vortex boundary travel around the vortex perimeter, which is
  consistent with the neutral stability reported in this regime in the
  present study. On the other hand, for fat vortex rings they observed
  that prolate perturbations of the shape of the vortex boundary are
  amplified and evolve into protrusions near the rear extremity of the
  vortex ring which then move upstream to become thin filaments.  This
  behavior too is consistent with the results of our analysis which
  predicts that fat vortex rings are linearly unstable to axisymmetric
  perturbations and in fact the deformation of the vortex boundary
  near the rear extremity of the vortex ring characterizing the most
  unstable eigenmodes, cf.~figure \ref{fig:uevec1}, can be viewed as a
  precursor of the filament observed at later stages of the evolution
  by \citet{Ofarrell_dabiri_2012}. Moreover, the value of the
  parameter $\alpha$ demarcating the two regimes reported by
  \citet{Ofarrell_dabiri_2012} was $\alpha = 0.7$, not too far from
  $\alpha_0 = 0.925$ found here.}

{We reiterate that} the physical significance of the results
reported here is {clearly} restricted by the assumption of
axisymmetry introduced to simplify the problem, which limits the class
of admissible perturbations. {More specifically, while the
  present analysis indicates that thin vortex rings are neutrally
  stable, they are in fact known to be linearly unstable and develop a
  bending instability when subject to perturbations depending on the
  azimuthal coordinate \citep[see Introduction for more
  discussion]{fukumoto_hattori_2005}.}  {In the same spirit, for
  Hill's vortex the studies of \citet{frk94,r99} demonstrated that
  perturbations with dependence on the azimuthal angle $\phi$ which
  are still localized near the rear stagnation point have in fact
  larger growth rates than the unstable modes with axial symmetry
  discussed by \cite{mm78,ProtasElcrat2016}.}  Thus, the ultimate goal
of this research program is to {address this limitation and}
provide a complete understanding of the stability of inviscid vortex
rings with respect to general perturbations; the present results are a
stepping stone in this direction.  This more general stability problem
can be studied using methods analogous to the approach developed in \S
\ref{sec:stab}, except that the perturbation $\rho$ will now also
depend on the azimuthal angle $\phi$ and the standard form of the
Biot-Savart kernel will need to be used in lieu of $\bK$ in
\eqref{eq:xeps2}--\eqref{eq:vne}.  Consequently, the eigenvalue
problem corresponding to \eqref{eq:evalp} will have both $\theta$ and
$\phi$ as independent variables. Solving this problem will be our next
step and results will be reported in the near future.

\section*{Acknowledgments}

The author acknowledges the support through an NSERC (Canada)
Discovery Grant.

\appendix 

\section{Evaluation of Improper Integrals $C_k(\theta)$ and $S_k(\theta)$}
\label{sec:CSk}

In this appendix we derive expressions \eqref{eq:Ck} and \eqref{eq:Sk}
for the improper integrals $C_k(\theta)$ and $S_k(\theta)$. Since
these integrals can be evaluated using standard techniques when $k=0$,
here we focus on the case when $k \ge 1$ and consider $C_k(\theta)$.
The integral is first transformed using standard trigonometric
identities as
\begin{align}
C_k(\theta) & = \int_0^{2\pi} \cos(k\theta') \ln\sin^2\left(\frac{\theta-\theta'}{2}\right) \, d\theta' \nonumber \\
& = \frac{1}{2} \int_0^{2\pi} e^{i k \theta'} \ln\left[\frac{e^{\frac{i (\theta - \theta')}{2}} - e^{-\frac{i (\theta - \theta')}{2}}}{2i} \right]^2 \, d\theta' +  
\frac{1}{2} \int_0^{2\pi} e^{-i k \theta'} \ln\left[\frac{e^{\frac{i (\theta - \theta')}{2}} - e^{-\frac{i (\theta - \theta')}{2}}}{2i}\right]^2 \, d\theta' \nonumber \\
& =: {\xi}_k(\theta) + \chi_k(\theta).
\label{eq:Ck2}
\end{align}
Since evidently $\chi_k(\theta) = \overline{{\xi}}_k(\theta)$, where
the overbar denotes complex conjugation, below we need to consider
only ${\xi}_k(\theta)$. This integral is mapped to the unit circle $C
:= \left\{ w \in \CC, \ |w| = 1 \right\}$ in the complex plane using
the substitutions $w = e^{\i \theta}$ and $-i dw = e^{i \theta}\,
d\theta$
\begin{align}
{\xi}_k(\theta) = - \frac{i}{2} \oint_C w'^{(k-1)} \ln \left[ -\frac{(w - w')^2}{4 w w'} \right] \, dw' 
\nonumber \\
= \frac{i}{2k} \oint_C \frac{w'^k + w'^{(k-1)} w }{w' - w} \, dw' = - \frac{\pi w^k}{k},
\label{eq:rhok} 
\end{align}
where the second equality follows from the integration by parts (since
$w, w' \in C$, the branch cut in the complex logarithm does not
intersect the unit circle so that this operation is allowed), whereas
the final result is obtained by applying Plemelj's identity to the
Cauchy-type singular integral \citep{Muskhelishvili2008}. Finally,
using \eqref{eq:Ck2} and returning to the original variable $\theta$,
we arrive at
\begin{equation}
C_k(\theta) = {\xi}_k(\theta) + \overline{{\xi}}_k(\theta) 
= - \frac{\pi w^k}{k} - \frac{\pi \overline{w}^k}{k} = - \frac{2 \pi}{k} \cos(k\theta).
\label{eq:Ck3}
\end{equation}
Identity \eqref{eq:Sk} for $S_k(\theta)$ is determined following
analogous steps.



\begin{thebibliography}{70}
\expandafter\ifx\csname natexlab\endcsname\relax\def\natexlab#1{#1}\fi

\bibitem[Akhmetov(2009)]{akhmetov-2009}
{\sc Akhmetov, D.~G.} 2009 {\em Vortex Rings\/}. Berlin, Heidelberg:
  Springer-Verlag.

\bibitem[Alekseenko {\em et~al.\/}(2007)Alekseenko, Kuibin \&
  Okulov]{alekseenko2007theory}
{\sc Alekseenko, S.V., Kuibin, P.A. \& Okulov, V.L.} 2007 {\em Theory of
  Concentrated Vortices: An Introduction\/}. Springer Berlin Heidelberg.

\bibitem[Arvidsson {\em et~al.\/}(2016)Arvidsson, Kov{\'a}cs, T{\"o}ger,
  Borgquist, Heiberg, Carlsson \& Arheden]{Arvidsson2016}
{\sc Arvidsson, Per~M., Kov{\'a}cs, S{\'a}ndor~J., T{\"o}ger, Johannes,
  Borgquist, Rasmus, Heiberg, Einar, Carlsson, Marcus \& Arheden, H{\aa}kan}
  2016 Vortex ring behavior provides the epigenetic blueprint for the human
  heart. {\em Scientific Reports\/} {\bf 6}, 22021 EP --, article.

\bibitem[Baker(1990)]{b90}
{\sc Baker, G.~R.} 1990 A study of the numerical stability of the method of
  contour dynamics. {\em Phil. Trans. Roy. Soc.\/} {\bf 333}, 391--400.

\bibitem[Berezovskii \& Kaplanskii(1992)]{bk92a}
{\sc Berezovskii, A.A. \& Kaplanskii, F.B} 1992 Dynamics of thin vortex rings
  in a low-viscosity fluid. {\em Fluid Dynamics\/} {\bf 27}, 643--649.

\bibitem[Boyd(2001)]{b01}
{\sc Boyd, J.~P.} 2001 {\em Chebyshev and Fourier Spectral Methods\/}. Dover.

\bibitem[Brooke~Benjamin(1975)]{Benjamin1975}
{\sc Brooke~Benjamin, T.} 1975 The alliance of practical and analytical
  insights into the nonlinear problems of fluid mechanics. In {\em Proc. Symp.
  on Applications of Methods of Functional Analysis to Problems in
  Mechanics\/}, , vol. 503. Springer.

\bibitem[Dabiri(2009)]{dabiri-2009-AR}
{\sc Dabiri, J.~O.} 2009 Optimal vortex formation as a unifying principle in
  biological propulsion. {\em Annual Review of Fluid Mechanics\/} {\bf 41}~(1),
  17--33.

\bibitem[Delfour \& Zol\'esio(2001)]{dz01a}
{\sc Delfour, M.~C. \& Zol\'esio, J.-P.} 2001 {\em Shape and Geometries ---
  Analysis, Differential Calculus and Optimization\/}. SIAM.

\bibitem[Dritschel(1985)]{d85}
{\sc Dritschel, D.~G.} 1985 The stability and energetics of corotating uniform
  vortices. {\em J. Fluid Mech.\/} {\bf 157}, 95--134.

\bibitem[Dritschel(1990)]{d90}
{\sc Dritschel, D.~G.} 1990 The stability of elliptical vortices in an external
  straining flow. {\em J. Fluid Mech.\/} {\bf 210}, 223--261.

\bibitem[Dritschel(1995)]{d95}
{\sc Dritschel, D.~G.} 1995 A general theory for two--dimensional vortex
  interactions. {\em J. Fluid Mech.\/} {\bf 293}, 269--303.

\bibitem[Dritschel \& Legras(1991)]{dl91}
{\sc Dritschel, D.~G. \& Legras, B.} 1991 {The elliptical models of
  two-dimensional vortex dynamics. II: Disturbance equations}. {\em Phys.
  Fluids A\/} {\bf 3}, 855--869.

\bibitem[Elcrat {\em et~al.\/}(2005)Elcrat, Fornberg \& Miller]{efm05}
{\sc Elcrat, A., Fornberg, B. \& Miller, K.} 2005 Stability of vortices in
  equilibrium with a cylinder. {\em J. Fluid Mech.\/} {\bf 544}, 53--68.

\bibitem[Elcrat \& Protas(2013)]{ep13}
{\sc Elcrat, A. \& Protas, B.} 2013 A framework for linear stability analysis
  of finite-area vortices. {\em Proceedings of the Royal Society A\/} {\bf
  469}, 20120709.

\bibitem[Fraenkel(1970)]{Fraenkel1970}
{\sc Fraenkel, L.~E.} 1970 On steady vortex rings of small cross-section in an
  ideal fluid. {\em Proceedings of the Royal Society of London A: Mathematical,
  Physical and Engineering Sciences\/} {\bf 316}~(1524), 29--62.

\bibitem[Fukumoto \& Hattori(2005)]{fukumoto_hattori_2005}
{\sc Fukumoto, Yasuhide \& Hattori, Yuji} 2005 Curvature instability of a
  vortex ring. {\em Journal of Fluid Mechanics\/} {\bf 526}, 77--115.

\bibitem[Fukumoto \& Kaplanski(2008)]{fukumoto-2008}
{\sc Fukumoto, Y. \& Kaplanski, F.~B.} 2008 Global time evolution of an
  axisymmetric vortex ring at low reynolds numbers. {\em Phys. Fluids\/} {\bf
  20}, 053103.

\bibitem[Fukumoto \& Moffatt(2000)]{fukumoto-2000}
{\sc Fukumoto, Y. \& Moffatt, H.~K.} 2000 Motion and expansion of a viscous
  vortex ring. part 1. a higher-order asymptotic formula for the velocity. {\em
  J. Fluid Mech.\/} {\bf 417}, 1--45.

\bibitem[Fukumoto \& Moffatt(2008)]{fukumoto-2008-PhysD}
{\sc Fukumoto, Y. \& Moffatt, H.~K.} 2008 Kinematic variational principle for
  motion of vortex rings. {\em Physica D\/} {\bf 237}, 2210--2217.

\bibitem[Fukuyu {\em et~al.\/}(1994)Fukuyu, Ruzi \& Kanai]{frk94}
{\sc Fukuyu, A., Ruzi, T. \& Kanai, A.} 1994 The response of {Hill's} vortex to
  a small three dimensional disturbance. {\em J. Phys. Soc. Japan\/} {\bf 63},
  510--527.

\bibitem[Gallay \& Smets(2018)]{gs18}
{\sc Gallay, Th. \& Smets, D.} 2018 Spectral stability of inviscid columnar
  vortices. ArXiv:1805.05064.

\bibitem[Giannuzzi {\em et~al.\/}(2016)Giannuzzi, Hargather \&
  Doig]{Giannuzzi2016}
{\sc Giannuzzi, P.~M., Hargather, M.~J. \& Doig, G.~C.} 2016 Explosive-driven
  shock wave and vortex ring interaction with a propane flame. {\em Shock
  Waves\/} {\bf 26}~(6), 851--857.

\bibitem[Golub(1973)]{Golub1973}
{\sc Golub, G.} 1973 Some modified matrix eigenvalue problems. {\em SIAM
  Review\/} {\bf 15}~(2), 318--334.

\bibitem[Graber(2015)]{vortex_gun}
{\sc Graber, Curtis~E.} 2015 Vortex cannon with enhanced ring vortex
  generation. US Patent US9217392B2.

\bibitem[Gumowski {\em et~al.\/}(2008)Gumowski, Miedzik, Goujon-Durand, Jenffer
  \& Wesfreid]{gmgdjw08}
{\sc Gumowski, K., Miedzik, J., Goujon-Durand, S., Jenffer, P. \& Wesfreid,
  J.~E.} 2008 Transition to a time-dependent state of fluid flow in the wake of
  a sphere. {\em Phys. Rev. E\/} {\bf 77}, 055308.

\bibitem[Guo {\em et~al.\/}(2004)Guo, Hallstrom \& Spirn]{ghs04}
{\sc Guo, Y., Hallstrom, Ch. \& Spirn, D.} 2004 {Dynamics Near an Unstable
  Kirchhoff Ellipse}. {\em Commun. Math. Phys.\/} {\bf 245}, 297--354.

\bibitem[Hackbusch(1995)]{h95}
{\sc Hackbusch, W.} 1995 {\em Integral Equations: Theory and Numerical
  Treatment\/}. Birkh\"auser.

\bibitem[Hattori \& Fukumoto(2003)]{HattoriFukumoto2003}
{\sc Hattori, Y. \& Fukumoto, Y.} 2003 Short-wavelength stability analysis of
  thin vortex rings. {\em Physics of Fluids\/} {\bf 15}~(10), 3151--3163.

\bibitem[Hattori \& Hijiya(2010)]{hh10}
{\sc Hattori, Y. \& Hijiya, K.} 2010 Short-wavelength stability analysis of
  {Hill's} vortex with/without swirl. {\em Phys Fluids\/} {\bf 22}, 074104.

\bibitem[Hicks(1899)]{Hicks1899}
{\sc Hicks, W.~M.} 1899 Ii. researches in vortex motion. {\textemdash}part iii.
  on spiral or gyrostatic vortex aggregates. {\em Philosophical Transactions of
  the Royal Society of London A: Mathematical, Physical and Engineering
  Sciences\/} {\bf 192}, 33--99.

\bibitem[Hill(1894)]{hill-1894}
{\sc Hill, M. J.~M.} 1894 On a spherical vortex. {\em Philos. Trans. Roy. Soc.
  London\/} {\bf A185}, 213--245.

\bibitem[Kamm(1987)]{k87}
{\sc Kamm, J.~R.} 1987 Shape and stability of two--dimensional vortex regions.
  PhD thesis, Caltech.

\bibitem[Kaplanski \& Rudi(2005)]{kaplanski-2005-PF}
{\sc Kaplanski, F.~B. \& Rudi, Y.~A.} 2005 A model for the formation of
  "optimal" vortex ring taking into account viscosity. {\em Phys. Fluids\/}
  {\bf 17}, 087101--087107.

\bibitem[Kelvin(1867)]{Kelvin1867}
{\sc Kelvin, Lord} 1867 The traslatory velocity of a circular vortex ring. {\em
  Phil. Mag.\/} {\bf 33}, 511–--512.

\bibitem[Kelvin(1880)]{k80}
{\sc Kelvin, Lord} 1880 Vibrations of a columnar vortex. {\em Phil. Mag.\/}
  {\bf 10}, 155--168.

\bibitem[Kheradvar \& Pedrizzetti(2012)]{kheradvar-2012}
{\sc Kheradvar, A. \& Pedrizzetti, G.} 2012 {\em Vortex Formation in the
  Cardiovascular System\/}. Springer.

\bibitem[Lamb(1932)]{lamb-1932}
{\sc Lamb, H.} 1932 {\em Hydrodynamics\/}. Dover, New York.

\bibitem[Laub(2005)]{l05}
{\sc Laub, A.~J.} 2005 {\em Matrix Analysis for Scientists and Engineers\/}.
  SIAM.

\bibitem[Lifschitz(1995)]{l95}
{\sc Lifschitz, A.} 1995 nstabilities of ideal fluids and related topics. {\em
  Z. Angew. Math. Mech.\/} {\bf 75}, 411.

\bibitem[Lifschitz \& Hameiri(1991)]{lh91}
{\sc Lifschitz, A. \& Hameiri, E.} 1991 Local stability conditions in fluid
  dynamics. {\em Phys. Fluids A\/} {\bf 3}, 2644--2651.

\bibitem[{Llewellyn Smith} \& Ford(2001)]{lsf01}
{\sc {Llewellyn Smith}, S.~G. \& Ford, R.} 2001 {Three-dimensional acoustic
  scattering by vortical flows. Part I: General theory}. {\em Phys. Fluids\/}
  {\bf 13}, 2876--2889.

\bibitem[Love(1893)]{l93}
{\sc Love, A. E.~H.} 1893 On the stability of certain vortex motions. {\em
  Proc. London Math. Soc.\/} {\bf s1--25}, 18--43.

\bibitem[Maxworthy(1977)]{maxworthy_1977}
{\sc Maxworthy, T.} 1977 Some experimental studies of vortex rings. {\em
  Journal of Fluid Mechanics\/} {\bf 81}~(3), 465--495.

\bibitem[Moffatt \& Moore(1978)]{mm78}
{\sc Moffatt, H.~K. \& Moore, D.~W.} 1978 The response of {Hill's} spherical
  vortex to a small axisymmetric disturbance. {\em J. Fluid Mech.\/} {\bf 87},
  749--760.

\bibitem[Mohseni(2001)]{mohseni-2001-PF}
{\sc Mohseni, K.} 2001 Statistical equilibrium theory for axisymmetric flow:
  Kelvin's variationalprinciple and an explanation for the vortex ring
  pinch--off process. {\em Phys. Fluids\/} {\bf 13}, 1924.

\bibitem[Moore \& Saffman(1975)]{MooreSaffman1975}
{\sc Moore, D.~W. \& Saffman, P.~G.} 1975 The instability of a straight vortex
  filament in a strain field. {\em Proceedings of the Royal Society of London
  A: Mathematical, Physical and Engineering Sciences\/} {\bf 346}~(1646),
  413--425.

\bibitem[Muskhelishvili(2008)]{Muskhelishvili2008}
{\sc Muskhelishvili, N.~I.} 2008 {\em Singular Integral Equations. Boundary
  Problems of Function Theory and Their Application to Mathematical Physics\/},
  2nd edn. Dover.

\bibitem[Norbury(1972)]{norbury-1972-proc}
{\sc Norbury, J.} 1972 A steady vortex ring close to {H}ill's spherical vortex.
  {\em Proc. Cambridge Phil. Soc.\/} {\bf 72}, 253--282.

\bibitem[Norbury(1973)]{norbury-1973-JFM}
{\sc Norbury, J.} 1973 A family of steady vortex rings. {\em J. Fluid Mech.\/}
  {\bf 57}, 417--431.

\bibitem[O'Farrell \& Dabiri(2012)]{Ofarrell_dabiri_2012}
{\sc O'Farrell, Clara \& Dabiri, John~O.} 2012 Perturbation response and
  pinch-off of vortex rings and dipoles. {\em Journal of Fluid Mechanics\/}
  {\bf 704}, 280--300.

\bibitem[Olver {\em et~al.\/}(2010)Olver, Lozier, Boisvert \& Clark]{olbc10}
{\sc Olver, F.~W.~J., Lozier, D.~W., Boisvert, R.~F. \& Clark, C.~W.}, ed. 2010
  {\em {NIST Handbook of Mathematical Functions}\/}. New York, NY: Cambridge
  University Press.

\bibitem[Pozrikidis(1986)]{p86}
{\sc Pozrikidis, C.} 1986 The nonlinear instability of {Hill's} vortex. {\em J.
  Fluid Mech.\/} {\bf 168}, 337 -- 367.

\bibitem[Protas \& Elcrat(2016)]{ProtasElcrat2016}
{\sc Protas, Bartosz \& Elcrat, Alan} 2016 {Linear stability of Hill's vortex
  to axisymmetric perturbations}. {\em Journal of Fluid Mechanics\/} {\bf 799},
  579--602.

\bibitem[Pullin(1992)]{p92}
{\sc Pullin, D.~I.} 1992 Contour dynamics methods. {\em Annual Review of Fluid
  Mechanics\/} {\bf 24}, 89--115.

\bibitem[Rozi(1999)]{r99}
{\sc Rozi, T.} 1999 {Evolution of the Surface of Hill's Vortex Subjected to a
  Small Three-Dimensional Disturbance for the Cases of $m=0,2,3$ and 4}. {\em
  J. Phys. Soc. Japan\/} {\bf 68}, 2940.

\bibitem[Rozi \& Fukumoto(2000)]{rf00}
{\sc Rozi, T. \& Fukumoto, Y.} 2000 {The Most Unstable Perturbation of
  Wave-Packet Form Inside Hill's Vortex}. {\em J. Phys. Soc. Japan\/} {\bf 69},
  2700--2701.

\bibitem[Saffman(1970)]{saffman-1970}
{\sc Saffman, P.~G.} 1970 The velocity of viscous vortex rings. {\em Studies in
  Applied Math.\/} {\bf 49}, 371.

\bibitem[Saffman(1992)]{saffman-1992}
{\sc Saffman, P.~G.} 1992 {\em Vortex Dynamics\/}. Cambridge, New York:
  Cambridge University Press.

\bibitem[Shariff {\em et~al.\/}(2008)Shariff, Leonard \& Ferziger]{slf08}
{\sc Shariff, K., Leonard, A. \& Ferziger, J.~H.} 2008 A contour dynamics
  algorithm for axisymmetric flow. {\em Journal of Computational Physics\/}
  {\bf 227}, 9044--9062.

\bibitem[Trefethen(2000)]{trefethen:SpecMthd}
{\sc Trefethen, L.~N.} 2000 {\em Spectral Methods in {Matlab}\/}. SIAM.

\bibitem[Trefethen(2013)]{n13}
{\sc Trefethen, N.} 2013 {\em Approximation Theory and Approximation
  Practice\/}. SIAM.

\bibitem[Tung \& Ting(1967)]{tung-1967}
{\sc Tung, C. \& Ting, L.} 1967 Motion and decay of a vortex ring. {\em Phys.
  Fluids\/} {\bf 10}, 901--910.

\bibitem[Wakelin \& Riley(1996)]{wr96}
{\sc Wakelin, S.~L. \& Riley, N.} 1996 Vortex ring interactions ii. inviscid
  models. {\em Quarterly Journal of Mechanics and Applied Mathematics\/} {\bf
  49}, 287--309.

\bibitem[Wan(1988)]{Wan1988}
{\sc Wan, Y.~H.} 1988 {Variational principles for Hill's spherical vortex and
  nearly spherical vortices}. {\em Trans. Amer. Math. Soc.\/} {\bf 308},
  299--312.

\bibitem[Widnall {\em et~al.\/}(1974)Widnall, Bliss \&
  Tsai]{widnall_bliss_tsai_1974}
{\sc Widnall, Sheila~E., Bliss, Donald~B. \& Tsai, Chon-Yin} 1974 The
  instability of short waves on a vortex ring. {\em Journal of Fluid
  Mechanics\/} {\bf 66}~(1), 35--47.

\bibitem[Widnall {\em et~al.\/}(1973)Widnall, Sullivan \&
  Owen]{WidnallSullivan1973}
{\sc Widnall, Sheila.~E., Sullivan, J.~P. \& Owen, Paul~Robert} 1973 On the
  stability of vortex rings. {\em Proceedings of the Royal Society of London.
  A. Mathematical and Physical Sciences\/} {\bf 332}~(1590), 335--353.

\bibitem[Widnall {\em et~al.\/}(1977)Widnall, yin Tsai \&
  Stuart]{WidnallTsaiStuart1977}
{\sc Widnall, Sheila.~E., yin Tsai, Chon \& Stuart, John~Trevor} 1977 The
  instability of the thin vortex ring of constant vorticity. {\em Philosophical
  Transactions of the Royal Society of London. Series A, Mathematical and
  Physical Sciences\/} {\bf 287}~(1344), 273--305.

\bibitem[Wu {\em et~al.\/}(2006)Wu, Ma \& Zhou]{wu-book-2006}
{\sc Wu, J.-Z., Ma, H.-Y. \& Zhou, M.-D.} 2006 {\em Vorticity and vortex
  dynamics\/}. Berlin, Heidelberg, New York: Springer-Verlag.

\bibitem[Ye \& Chu(1995)]{YeChu1995}
{\sc Ye, Qu-Yuan \& Chu, C.~K.} 1995 Unsteady evolutions of vortex rings. {\em
  Physics of Fluids\/} {\bf 7}~(4), 795--801.

\end{thebibliography}

\end{document}